# Optimizing high-temperature electron mobility in single-crystal $Bi_2O_2Se$ based on its unconventional dependence on concentration.

Antonín Sojka[1], Petr Knotek[1], Jan Zich[1], Martin Míšek[2], Roman Tesař[2], Kyo-Hoon Ahn[2], Petr Levinský[2], Jiří Navrátil[1], Pavlína Ruleová[1], Jiří Hejtmánek[2], Karel Knížek[2], Václav Holý[3,4], and Čestmír Drašar[1,*]

[1]*University of Pardubice, Faculty of Chemical Technology, Studentská 573,53210 Pardubice, Czech Republic*
[2]*Institute of Physics of the Czech Academy of Sciences, Cukrovarnická 10/112,162 00 Prague 6, Czech Republic*
[3]*Department of Condensed Matter Physics, Faculty of Mathematics and Physics, Charles University, Ke Karlovu 3, 121 16 Praha 2, Czech Republic*
[4]*Masaryk University, Department of Condensed Matter Physics, Kotlářská 2, 61137 Brno, Czech Republic*

**ABSTRACT**. Quasi-2D $Bi_2O_2Se$ is part of an intensive materials research effort aimed at finding new semiconductors that outperform silicon-based electronics in terms of speed and power consumption. This material exhibits exceptionally high carrier mobility at low temperatures but mediocre mobility at 300 K. Its high mobility is generally associated with its high permittivity ($\varepsilon_r$≈500), which is also associated with metallicity persisting down to very low carrier concentrations. This material exhibits a counterintuitive increase in carrier mobility as the concentration increases. The connection between low both carrier concentration and mobility in Se-rich conditions, and both high carrier concentration and mobility in Se-poor conditions suggests that the increase is related to native defects. We demonstrate that these defects can alter the effective mass of charge carriers. Specifically, substitutional $Se_{Bi}$ defects, which appear under Se-rich conditions, destroy the $Bi_2O_2$ channel and compromise charge transport properties. These defects increase the effective mass of charge carriers transforming the original semiconductor into a semimetal and introducing holes into charge transport. Additionally, we show that single crystals are generally inhomogeneous, particularly those grown under Se-rich conditions. Unlike Se-poor conditions, Se-rich conditions induce a higher concentration of dislocations and extraneous phases. These findings suggest that the perfection of the $Bi_2O_2$ channel is crucial for mobility, particularly at room temperature.

## I. INTRODUCTION.

The pursuit of novel high-speed, low-consumption electronics is largely focused on 2D materials that naturally facilitate formation of low-dimensional structures [1–4]. Quasi-2D $Bi_2O_2Se$ is one of the most promising materials [5–8] outperforming both silicon-based electronics and other 2D materials. This material's extraordinary charge carrier mobility, intrinsic insulating layer ($Bi_2SeO_5$) and perhaps also stability in air have motivated the preparation of low-voltage field-effect transistors based on $Bi_2O_2Se$ [9]. However, the correlation between the structure, defects, and physical properties is still debated. According to ARPES and optical and transport experiments, it is a semiconductor with 0.8 – 0.9 eV band gap [10–13]. This contradicts the observed metallicity appearing down to very low electron concentrations (≈$10^{16}$ $cm^{-3}$ [14]; only ≈$10^{14}$ $cm^{-3}$ according to our experiments.) The observed metallicity has two preconditions. The very high permittivity ($\varepsilon_r$≈500) and the low effective mass of conduction electrons [12,15] resulting in large effective Bohr radius [14]. The sporadic observation of hidden direct band gap illustrates the complexity of this material [11]. In addition, the reported effective mass ranges from $0.032m_0$ [16] to $0.23m_0$ [15], suggesting that the defect structure may play an important role.

Carrier mobility $\mu_H$ as high as 470000 $cm^2V^{-1}s^{-1}$ [17] and 280000 $cm^2V^{-1}s^{-1}$ [10] at 2 K, and 100 000 $cm^2V^{-1}s^{-1}$ and 900 $cm^2V^{-1}s^{-1}$ at 2K and 300 K in one of the present samples with carrier concentration n≈$10^{17}$ $cm^{-3}$ can be called ultrahigh; with carrier effective mass $m^* = 0.13\ m_e$, this gives long relaxation times $\tau \approx 3 \cdot 10^{-10} s$ and $\approx 5 \cdot 10^{-13} s$, respectively. Such $\mu_H$ values exceed the theoretical predictions ($\mu_H$≈100 000 $cm^2V^{-1}s^{-1}$ and ≈210 $cm^2V^{-1}s^{-1}$, respectively for relatively low carrier concentration N=$10^{16}$ $cm^{-3}$) [18]. Several points contributing to such high mobility can be found in the literature. The first point is the so-called self-modulation doping [19], which suggests that the Se layer is a doping part of the structure due to the presence of Se vacancies, $V_{Se}$, while the $Bi_2O_2$ layer is a defect-free conduction channel. Second point is the very high out-of-plane permittivity $\varepsilon_r \approx 170$ [14] and in-plane permittivity $\varepsilon_r \approx 500$ [12] which significantly reduces electrostatic interactions between dopant species and itinerant electrons thus increasing their mobility. Both together provide a solid basis for the high mobility in low temperature range. None of the papers analyze the anomalous dependence of carrier mobility on carrier concentration. Carrier mobility increases with carrier concentration [14,15], despite the fact, that phonon scattering of charge carriers is proportional to their concentration [20]. This increase is notable for low temperature mobility but is still observable for room-temperature mobility.

In addition, there are other peculiarities mentioned in the literature: the T-square resistivity discussed in the literature lacks an explanation [15]. In some samples, 2D transport features are observed [16]. Ref. [21] discusses the transition of the paraelectric $Bi_2O_2Se$ to ferroelectric phase due to the transition from a tetragonal to an orthorhombic structure. Similarly, authors of [18] suggest that biaxial deformation of the lattice induces high permittivity and "giant" modulation of electron mobility. Clearly, any small piece of the puzzle can help improve our understanding of this material.

This work focuses on high-temperature mobility limits and the counterintuitive step-like increase in Hall mobility with carrier concentration. The starting point is the extraordinary permittivity and phononic spectrum of $Bi_2O_2Se$, which enhances the carrier mobility in the low-temperature region by screening of charged defects. However, this alone does not explain the unconventional increase of mobility of the carriers with their concentration and variations in the high-temperature carrier mobility in the literature. Both the literature data and our data suggest that native defects Se vacancies (Se-poor conditions) scatter free electrons less effectively than $Se_{Bi}$ (Se-rich conditions) do. While this could explain the differences in mobility at low temperatures, it cannot account for the increase in mobility with concentration or high-temperature mobility itself. Therefore, additional factors must be considered in assessment of the carrier mobility and crystal quality.

Native defects may affect the effective mass of free carriers. For instance, $V_{Se}$ and $O_{Se}$ defects, which are associated with Se-poor conditions, have a negligible impact on the in-plane effective mass as indicated by Density Functional Theory (DFT) calculations. In contrast, $Se_{Bi}$ defects, which are associated with Se-rich conditions, result in a significant increase in the in-plane effective mass. In addition, Se-rich conditions result in a heterogeneous distribution of native defects, which can lead to the formation of Fermi lakes rather than a Fermi sea in the low concentration region [15,22,23]. Furthermore, $Se_{Bi}$ defects alter the nature of the band structure from semiconducting to semimetallic, resulting in multiple types of charge carriers. This could make the Hall mobility analysis ambiguous (see Supplementary material (SM) [24], section 1 and 8). The nature and concentration of point defects are associated with the crystals ´stoichiometry and, consequently, with the concentration of dislocations and eventually extraneous phases in a "single crystal" sample, if any. In addition to stoichiometry, the growth temperature and the temperature gradient play an important role.

Although our experiments do not provide definitive evidence, we believe that our findings could be useful in optimizing this material for room-temperature applications.

## II. METHODS

### A. Synthesis and crystal growth

The $Bi_2O_2Se$ single crystals were grown via a gas-phase transport reaction in a temperature gradient. The starting materials, high purity $Bi_2O_3$ (5N powder) (~1% overstoichiometry of $Bi_2O_3$), and $Bi_2Se_3$ (synthesized from Bi and Se of 5 N purity) all chemicals from Merck, were weighed into quartz glass ampoules. Single crystals were grown over the course of one month using a horizontal furnace with a gradient of 2 $Kcm^{-1}$. Sample A was grown at 840 - 850 °C, followed by slow stepwise cooling 100 °C per week until reaching room temperature. Sample B was grown at 810-820 °C and cooled in an off furnace. This was followed by annealing the sample for 1 week at 600°C, then quenching it to room temperature in ambient air.

Other $Bi_2O_2Se$ samples (C, D, E and F and doped samples) were synthesized using a low-temperature solid-state route. High-purity bismuth (5N, Sigma-Aldrich) was first pulverized in an MM500 nano oscillating mill operating at 30 Hz for 20 min. Elemental selenium (5N, Sigma-Aldrich) was finely ground in an agate mortar, while $Bi_2O_3$ powder (5N, Alfa Aesar) was pre-treated by calcination in flowing argon at 450 °C for 30 min to remove carbonate impurities, followed by rapid cooling. The obtained powders were mixed in a stoichiometric ratio in an agate mortar and sealed in evacuated quartz ampoules (residual pressure < $10^{-3}$ Pa). The ampoules were initially heated at 0.1 K $min^{-1}$ to 300 °C and kept for 24 h, then ramped at 1 K $min^{-1}$ to 400 °C and maintained for 10 days while being continuously rotated (~15 rpm) to ensure reaction uniformity. After natural cooling, the precursor ($Bi_2O_2Se$) still in sealed ampoules were transferred to a gradient furnace. For crystal growth, the temperature was increased at 1 K $min^{-1}$ to a temperature gradient 840 - 850 °C, where it was held for three weeks. Finally, the ampoules were cooled slowly to room temperature at 0.1 K $min^{-1}$.
The final products of the synthesis/growth were well crystallized $Bi_2O_2Se$ with the crystal ranging in size from approximately 10 $\mu$m to 10 mm. The larger, well-grown $Bi_2O_2Se$ single crystals with sizes ranging from approximately 4×4 to 7×7 mm were hand-picked and used for experiments.

### B. XRD techniques

X-ray diffraction measurements were carried out on a RIGAKU-Smartlab rotating anode diffractometer (45kV/200mA) using $CuK\alpha_1$ radiation. The diffractometer was equipped with a primary parabolic multilayer mirror and a 2x220Ge channel-cut monochromator, we used a two-dimensional x-ray detector in one-dimensional counting mode. We measured long symmetric 2θ/ω scans for diffraction angles 2θ from 10deg up to 100deg, as well as symmetric and asymmetric reciprocal-space maps around the reciprocal lattice points 0 0 6 and 2 0 10, respectively. The measured scans and maps are displayed in the SM [24].
From the positions of the diffraction maxima in the 2θ/ω scans we determined the vertical lattice parameter c using the Cohen-Wagner plot [25]. The vertical thickness of coherent domains was estimated from the widths of the diffraction maxima by means of the Williamson-Hall method [26]. The heights of the diffraction maxima were fitted to a structure model, in which the density of Se atoms on Bi lattice places was fitted; from the fit we also determined the distance between the Bi and Se (001) planes. For the fit we used a semikinematical model [27]. We integrated the intensities of the symmetric 0 0 6 reciprocal space maps along the 2θ/ω axis and from the width of the resulting curves we estimated mean angular misalignment of the mosaic blocks, from the asymmetric 2 0 10 reciprocal-space maps we determined the lattice parameter *a*.

### C. Atomic force microscopy (AFM) techniques

AFM measurements were performed in ambient air in two different modes immediately after cleaving single crystal surfaces. 1) PeakForce QNM (Quantitative Nanomechanical Mapping) mode, which is used to simultaneously acquire topography and map mechanical properties (evaluated as the logarithm of the DMT modulus) and adhesion (see refs. [28,29] using an AFM Dimension ICON (Bruker); and 2) two-pass Kelvin probe force microscopy (KPFM), which is used to determine topography and surface potential (see refs. [30,31] using an AFM Dimension Nexus (Bruker).

### D. Density functional theory (DFT) calculations

Calculations of structures with an introduced defect were performed in a 3×3×1 supercell with resulting lattice parameters of 11.673×11.673×12.213 Å and containing a total of 90 atoms. The size of the supercell ensures a similar defect spacing in all three directions and represents a reasonable compromise between computational accuracy and calculation time requirements. The calculations were made with the WIEN2k program [32]. This program is based on the Density Functional Theory (DFT) and uses the full-potential linearized augmented plane-wave (FP LAPW) method with the dual basis set. In the LAPW methods, space is partitioned into atomic spheres and the interstitial region. Electron states are then divided into core states, which are fully contained within the atomic spheres, and valence states, which can be delocalized among the spheres and the interstitial region. The radii of the atomic spheres were taken to be 2.325 a.u. for Bi, 1.86 a.u. for O, and 2.325 a.u. for Se. We used a plane wave cutoff described by $R_{mt}K_{max}$ = 7.5. The number of k-points in the irreducible part of the Brillouin zone was 172, which is sufficient for the large supercell used. The positions of all atoms were relaxed with respect to minimal calculated forces for each type of defect. Since we expected spin-polarized solutions in some cases, all calculations were performed as spin-polarized for the sake of comparison. We used the modified Becke–Johnson (mBJ) potential, which is appropriate when the bands around the Fermi surface are predominantly of the s and p character [33]. This potential provides a better description of the band gaps, in particular. However, since mBJ potential is not suitable for calculating forces, it cannot be used for optimization of atom positions. For this purpose and for comparative calculation of formation energy, we also used the Generalized Gradient Approximation (GGA) potential [34]. Effective masses were calculated using the perturbative ***k.p*** approach implemented in the WIEN2k code, which allows accurate determination of the band curvature by evaluating the second derivatives of the energy with respect to the crystal momentum [35].

### E. Transport measurements

Electrical resistivity $\rho$ and Hall coefficient $R_H$ were measured between 300 K and 2 K using the Electrical Transport Option (ETO) of the Quantum Design Physical Property Measurement System (PPMS) on samples about 0.3 mm thick. The four-probe method was used with silver leads attached to the samples with Ag paint. Magnetoresistance in fields up to ±14 T was measured at selected temperatures with magnetic field perpendicular to electrical current. Hall carrier concentration $n_H$ was calculated from $R_H$ measured in magnetic field ±0.8 T using the relation $n_H = 1/(e\ R_H)$.

## III. RESULTS AND DISCUSSIONS

The present study is based on the comparison of two different samples belonging to two groups with presumably different defect structures. One group (A) of samples is Se-rich dominated by selenium atom on the bismuth site, $Se_{Bi}$ and one (B) is Se-poor dominated by selenium vacancies $V_{Se}$ (see section 1 in SM [24]). In line with literature, we find that the measurement of the SdH effect is difficult to impossible in the Se rich samples, while the Se-poor sample shows nice oscillations (see e.g. [15,16]). The electron concentration and mobility and the residual resistivity ratio (RRR) can be used as hallmarks for Se-rich/Se-poor composition with the borderline $n_b$ ($T \approx 300K$) ≈$10^{17}$ cm$^{-3}$, $\mu_H$(T ≈ 2K) ≈ $5 \times 10^4$cm$^2$V$^{-1}$s$^{-1}$ and RRR≈100 [14,15]. Figure 1a and Table I summarize the conclusions drawn from the literature that are consistent with our findings.

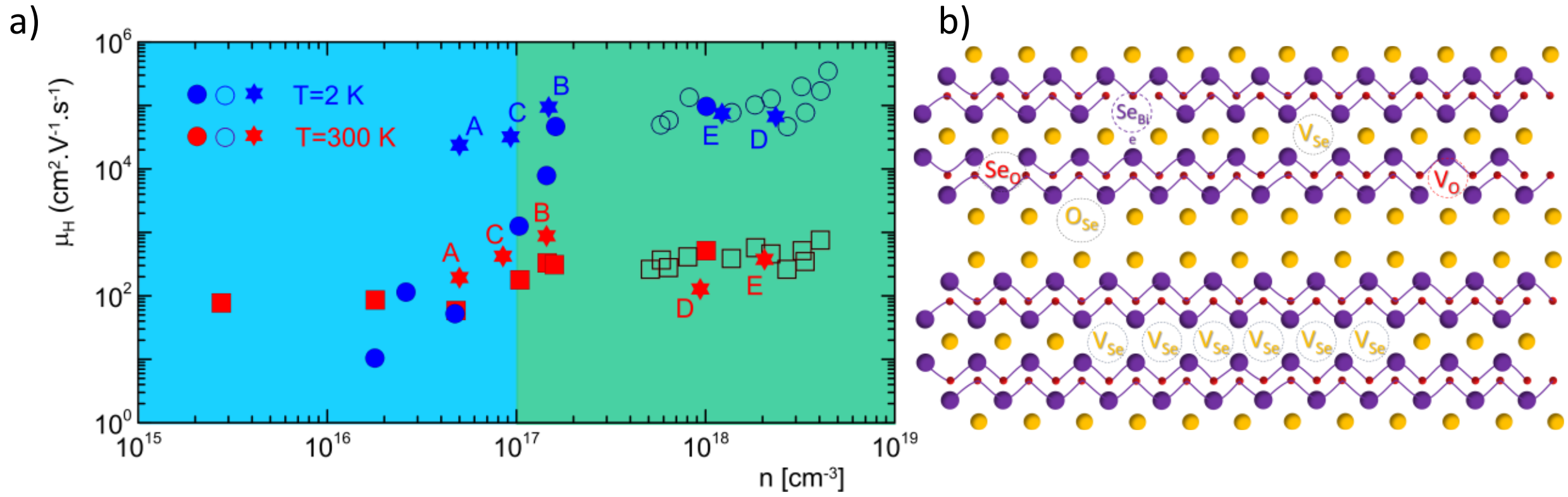

FIG 1. a) Carrier mobility as a function of electron concentration $\mu_H = f(n)$. An increase in mobility with concentration completely contradicts the power law typically observed in semiconductors [22]. 1) Solid symbols (squares and circles) are adopted from Ref. [14] and the open symbols are adopted from Ref. [15]. Samples A and B are close to the borderline and are examined closely in this study (asterisks). These samples exhibit a similar trend but with higher mobility than the published data. This difference can be attributed to the quality and size of the single-crystalline samples. In contrast to the cited works have used large (≈5 mm) and thick (≈0.3 mm) samples. b) The structure of $Bi_2O_2Se$ with single point defects $V_{Se}$, $V_O$, $Se_O$, $O_{Se}$ and $Se_{Bi}$ (top) and a multiple point defect n $V_{Se}$ (bottom). The transport properties of samples C, D and E are presented in section 8 of SM [24].

TABLE I. Properties of the two groups of $Bi_2O_2Se$ crystals analyzed in this study. Group A is represented by sample A, and group B by sample B.

| Parameter | Group A | Group B |
|---|---|---|
| Composition /stoichiometry | Se rich | Se poor |
| Main point defects | $Se_{Bi}$ | $V_{Se}$ |
| Charge transport (assumed) | Diluted semimetal | Diluted metal |
| Hall mobility (2K) | Low <5 m$^2$V$^{-1}$s$^{-1}$ | High >5 m$^2$V$^{-1}$s$^{-1}$ |
| Hall n | low<$10^{17}$ cm$^{-3}$ | high>$10^{17}$ cm$^{-3}$ |
| SdH | - | observable |
| RRR (2K/300K) | low<100 | high>100 |

To study the increase in Hall mobility with carrier concentration, we need to reveal any differences between the two samples A and B in structure, band structure, and physical properties.

### A. Structure and composition – high-resolution XRD and EDS

The EDS (energy dispersive spectroscopy) analysis (SM [24], sec. 2) indicates that both samples are slightly off-stoichiometric, with one sample being Se-poor and the other Se-rich. The average results give the stoichiometry $Bi_{1.997}O_{1.998}$ $Se_{1.005}$ (±0.01. 0.02. 0.01. respectively) for sample A and $Bi_{1.991}O_{2.014}$ $Se_{0.995}$ (±0.04. 0.06. 0.02.) for sample B. However, this is inconsistent with the high resolution (HR)XRD analysis [36]. Analysis of the HRXRD experiments suggested the presence of $Se_{Bi}$ in both samples A and B with a composition close to the formula $Bi_{1.9}O_2Se_{1.1}$ (see SM [24], Fig. 3d). This suggests a reduced Bi site occupation factor. While this analysis is not absolute, it shows the most plausible scenario. In other words, the fitting procedure does not match any other point defect. Thus, the HRXRD data only indicates a reduced x-ray scattering at the Bi site. Note that the concentration of charge carriers and thus the corresponding point defects is about three orders of magnitude lower than the suggested EDS non-stoichiometry. The presence of $V_{Se}$ and $Se_{Bi}$ is more or less consistent with our and published DFT data; compare Table S5 [24] and Refs. [19,37–39]. However, the published formation energies of native defects based on DFT are generally very inconsistent. The exception is $V_{Se}$, which consistently has one of the lowest formation energies. In contrast with literature, our calculations show that the native defect with the lowest formation energy is $O_{Se}$, oxygen on the selenium site. Although this defect has little influence on electron concentration or band structure (see Figure 2, Table II and SM [24], section 6), we include it in our analysis because it reflects a shift in stoichiometry.

### B. Band structure

In addition to calculating the formation energies of defects (see section 5 of SM [24]), we theoretically investigated how the point defects influence the band structure and the effective masses as an important mobility parameter. Due to the peculiarities mentioned in the introduction, we used two potentials: mJB (designed for s-p compounds) and GGA (which underestimates the band gap), to better understand any tendencies. Of all the defects studied, Se vacancies, ($V_{Se}$) and Se antisites ($Se_{Bi}$) had the most significant impact. According to both potentials, $V_{Se}$ defects induce a significant electron concentration, shifting the Fermi level into the conduction band. $Se_{Bi}$ defects are the only ones that induce a distinct defect-based band within the band gap eventually resulting in a crossover from a semiconductor to a semimetal behavior involving both electrons and holes close to the Fermi level (see Figure 2). The observation of a hidden direct band gap in an ARPES experiment may be related to this phenomenon [11]. This aligns with the conclusion drawn from THz cyclotron resonance (see SM [24], section 4), which shows the presence of multiple resonances (i.e. carrier types) in sample A and a single resonance in sample B, indicating a single type of electron. These results suggest that sample A is dominated by $Se_{Bi}$, and that sample B is dominated by $V_{Se}$. However, both defects are likely to be present in both samples. $O_{Se}$, which has the lowest formation energy, has a negligible influence on the band structure and is likely present to balance the stoichiometry (see sections A above and 2,5, and 6 of SM [24]).

The effective masses of charge carriers in the high-symmetry directions were obtained from band structure calculations (see Figure 2). Due to possible doping by multiple defects, the precise filling of particular bands is unknown. For this reason, we compare the effective masses at different Fermi levels. Therefore, the data are only indicative, not conclusive. The most significant data are summarized in Table II and SM [24], section 6. The data suggests that $Se_{Bi}$ defects induce greater effective masses than a defect-free (perfect) $Bi_2O_2Se$ structure. Additionally, these defects shift the CBM from the Γ point to the M point (Table II). At the GGA potential, these defects induce semimetallic state, meaning the presence of both electrons and holes. Therefore, the Hall concentration/mobility analysis can lead to an under- or overestimation of carrier mobility (see SM [24], section 1).

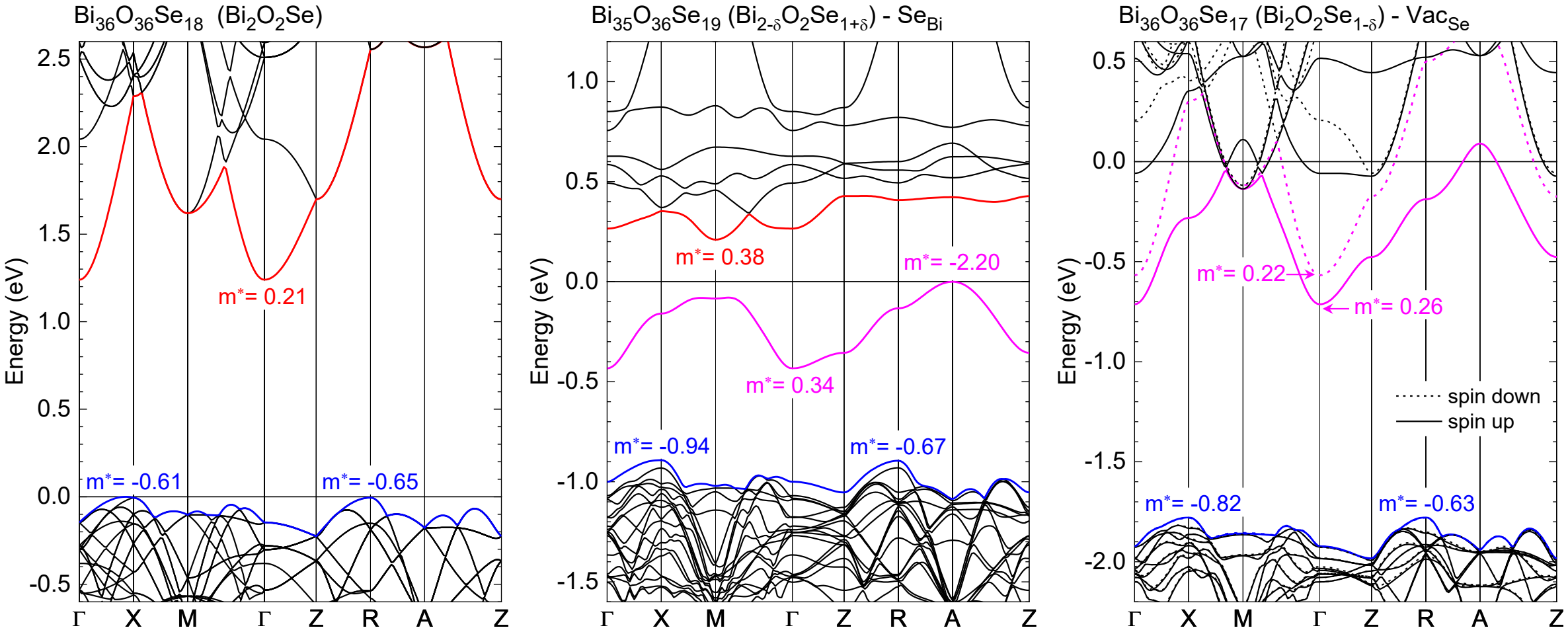

FIG 2. Modification of the band structure and electron effective mass m* due to the occurrence of $V_{Se}$ and $Se_{Bi}$ defects, calculated for mBJ potential. GGA underestimates the width of the band gap ($E_g$) whereas mBJ overestimates it (ARPES-derived $E_g \approx 0.8$ eV). Note that mBJ is designed to be more suitable for compounds with s- and p-bonding.

TABLE II. Changes in effective mass of charge carriers m*, induced by native defects in $Bi_2O_2Se$. The effective masses are calculated for the bands closest to the Fermi level ($E_F$) for selected symmetry points. The position of CBM against $E_F$ is negative when the band is merged into Fermi sea. Average of m* is calculated as weighted by Fermi-Dirac distribution for 300 K. Band position (BP) is relative to the Fermi level - **above $E_F$** - 1 (red line in Figure. 2), **around** $E_F$ - 2(magenta line in Figure. 2) or **below** $E_F$ - 3 (blue line in Figure. 2).

| | Γ | X | M | Z | A | average | BP |
|---|---|---|---|---|---|---|---|
| $Bi_{36}O_{36}Se_{18}$ / - | 0.20 | 0.49 | 0.22 | 0.34 | 0.56 | 0.21 | 1 |
| | 0.19 | -0.84 | -2.36 | 0.13 | -2.54 | -0.88 | 3 |
| $Bi_{35}O_{36}Se_{19}$/ $Se_{Bi}$ | 0.78 | -0.47 | 0.36 | -2.19 | 2.15 | 0.39 | 1 |
| | 0.34 | 2.80 | 1.01 | 0.57 | -2.28 | -2.76 | 2 |
| | 0.42 | -0.94 | 0.45 | 0.28 | 0.14 | -0.88 | 3 |
| $Bi_{36}O_{36}Se_{17}$ / $V_{Se}$ | 0.22/0.25 | -0.09/2.38 | 0.24/0.25 | -0.56/0.98 | -0.12/-0.38 | 0.23/0.33 | 2 |
| spin up / down | 0.15/0.16 | -0.83/-0.81 | 11.08/12.33 | 0.12/0.12 | 0.05/0.05 | -0.87/-0.86 | 3 |
| $Bi_{36}O_{37}Se_{17}$ / $O_{Se}$ | 0.21 | -0.02 | 0.23 | -0.21 | -0.04 | 0.22 | 1 |
| | 0.74 | -12.01 | 1.19 | 0.69 | 0.85 | 1.95 | 3 |

The effective mass of charge carriers increases due to both $V_{Se}$ and $Se_{Bi}$ defects, but $Se_{Bi}$ has a more significant effect. The lighter carriers may be those observed in SdH (see [15,40]) and Section C below). Therefore, we can conclude that both defects, $V_{Se}$ and $Se_{Bi}$, compromise charge transport in terms of effective mass; however, this parameter is less compromised for $V_{Se}$. This is consistent with the reported effective mass of $m^*$= 0.032 in highly defective Se-poor samples [16]. Additionally, $V_{Se}$ is less accessible to scattering processes due to its position in the energy spectrum [19]. The analysis of the effective mass of charge carriers shows that the nature of native defects can profoundly affect their effective mass. Thus, a non-negligible concentration of defects may induce very different carrier effective masses at Fermi level. In terms of carrier masses or mobility, we can therefore have both "proper" and "improper" defects, reflecting "improper" or "proper" stoichiometry of the crystal. Additionally, improper stoichiometry (i.e., Se-rich) can induce a heterogeneous structure associated with a compromised mobility as discussed in section D.

Regarding heterogeneity, we can either assume an inhomogeneous distribution of the native defects (Figure 1 b), or the presence of 2D inclusions of extraneous phases embedded in the 2D $Bi_2O_2Se$ matrix. Heterogeneous doping has been suggested in references [14,16]. The authors of ref. [15] proposed a heterogeneous doping as a means of explaining the difference in Hall and SdH concentrations. This distribution of defects is also consistent with the broken zipper regions observed in ref. [41]. We believe that a "improper" stoichiometry not only causes a heterogeneous distribution of defects, but is also associated with larger defects such as dislocations, grain boundaries, and tiny 2D particles of related extraneous phases ($Bi_2SeO_5$, BiSe, $Bi_2Se_3$…) that play a significant role in mobility. However, the effect of such macroscopic defects could be twofold. They can either serve as doping centers leaving the rest of the structure defect-free and forming high-mobility channels (i.e., modulation doping), or they can be significant scattering centers. The AFM techniques suggest the latter.

### D. AFM: investigating the topology, mechanical properties and electrical properties of the $Bi_2O_2Se$

Figure 3 compares samples A and B, which were used for the transport measurements. These samples repeatedly exhibited significant differences in their freshly cleaved surfaces in AFM images. Sample B is nearly flawless, while sample A contains three types of defects:

a) rectangular lamellae with heights of ≈1.1 nm and lateral dimensions on the order of micrometers. Their thickness suggests a single $Bi_2SeO_5$ lamella (with a lattice parameter c ≈1.14 nm) (Figures 3 A and B). This is consistent with the Kelvin probe scan showing a distinct work function compared to the surrounding surface, which is presumably $Bi_2O_2Se$ (Figures 3D and 3E). However, this is inconsistent with the mechanical and adhesion responses of these lamellae, which are identical to those of the surrounding surface (see Figure 4 D, E, G and H as well as Figure S10). The HRXRD of $Bi_2O_2Se$ sample A shows an extraordinary reflection at 2Θ ≈18°, which is attributable to $Bi_2SeO_5$ (see Figure S3a [24]). Additionally, only the theoretically predicted Raman modes of $Bi_2O_2Se$ were detected in the Raman spectra of both samples, which were taken from large crystal surfaces. However, other phonons corresponding to the $Bi_2SeO_5$ phase were detected at the edges of the sample A (not shown).

b) nonrectangular lamellae with heights of ≈1.3 nm and lateral dimensions on the order of micrometers can be identified as a single $Bi_2O_2Se$ lamella (with a lattice parameter c ≈1.22 nm) (see Figures 3A and 3B). This is consistent with the Kelvin probe scan showing the same work function as the surrounding surface (Figure 3D and 3E). However, it is inconsistent with mechanical behavior. We observe higher stiffness, represented by the orange–red contrast in Figures 4D and 4E and reduced adhesion represented by the dark red contrast in Figures 4G and 4H; Although the difference in height between the two defects is clearly visible, it lies within the measurement error. Therefore, this could be interpreted differently; for example, as two $Bi_2O_2Se$ with different defect structure or as fragments of 2D phases from the Bi-Se system.

c) irregularly distributed protrusions ≈3-4 nm (up to 10 nm) in height. These protrusions are characterized by significantly higher stiffness and lower adhesion. Note that the protrusions are distributed very inhomogeneously (see Figures 4A and 4B and Figure S11 [24]).

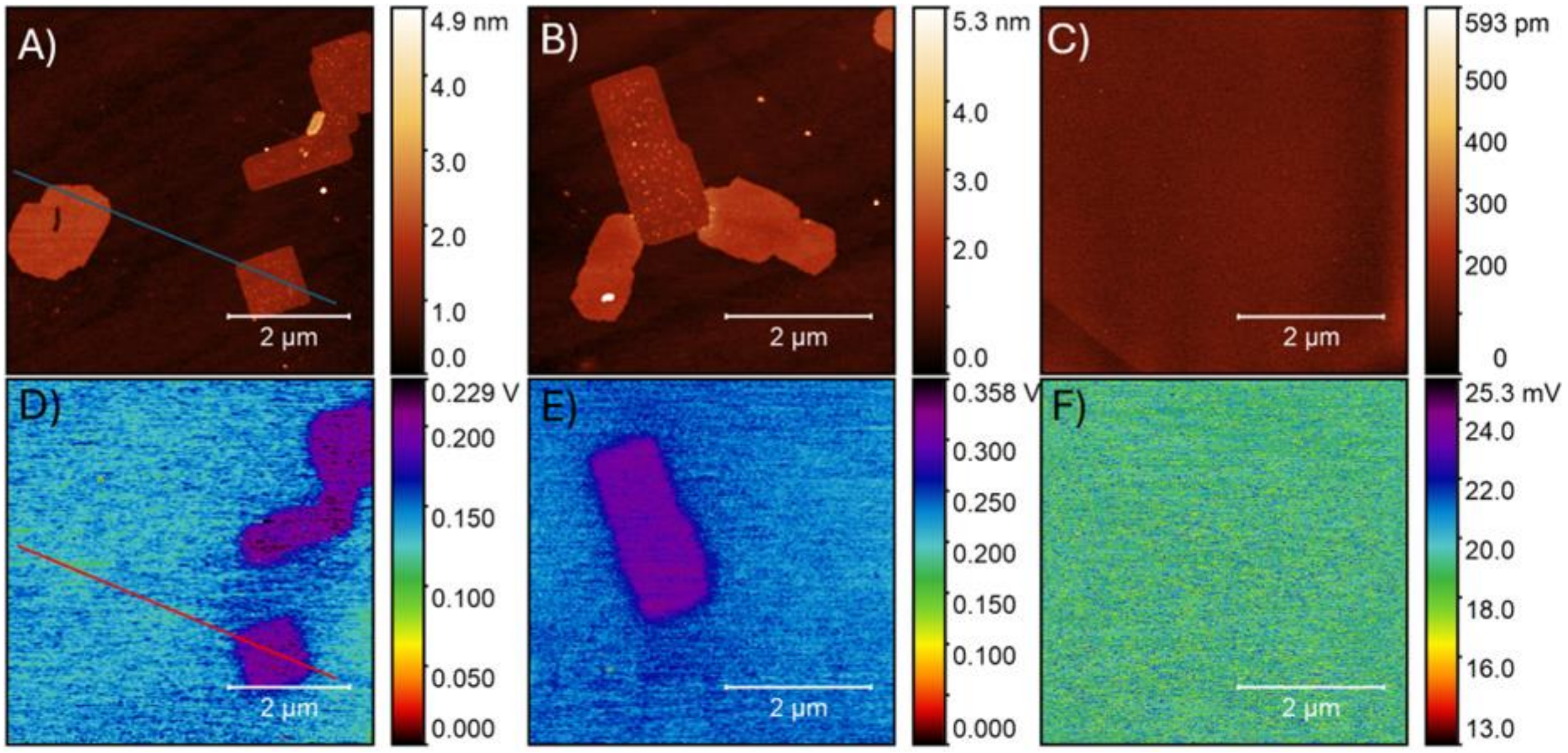

FIG 3. AFM Kelvin probe force microscopy (KPFM) images of the freshly cleaved surfaces. Topography maps of sample A, (Figures A, and B) and sample B (Figure C) compared with the corresponding maps of surface potential reflecting the relative changes in work function (D-F). In sample A, rectangular lamellae detected in the topography showed a surface potential increased by approximately 100 mV compared to their surroundings. In contrast, the non-rectangular lamellae were not detectable in the surface-potential images. The protrusions were not detected due to the lower lateral resolution typical to the two-pass KPFM technique [30]. The measurements were reproduced many times with similar output. Topography and surface potential line profiles indicated by line markers in Figures A and D are presented in detail in Figure S13 [24].

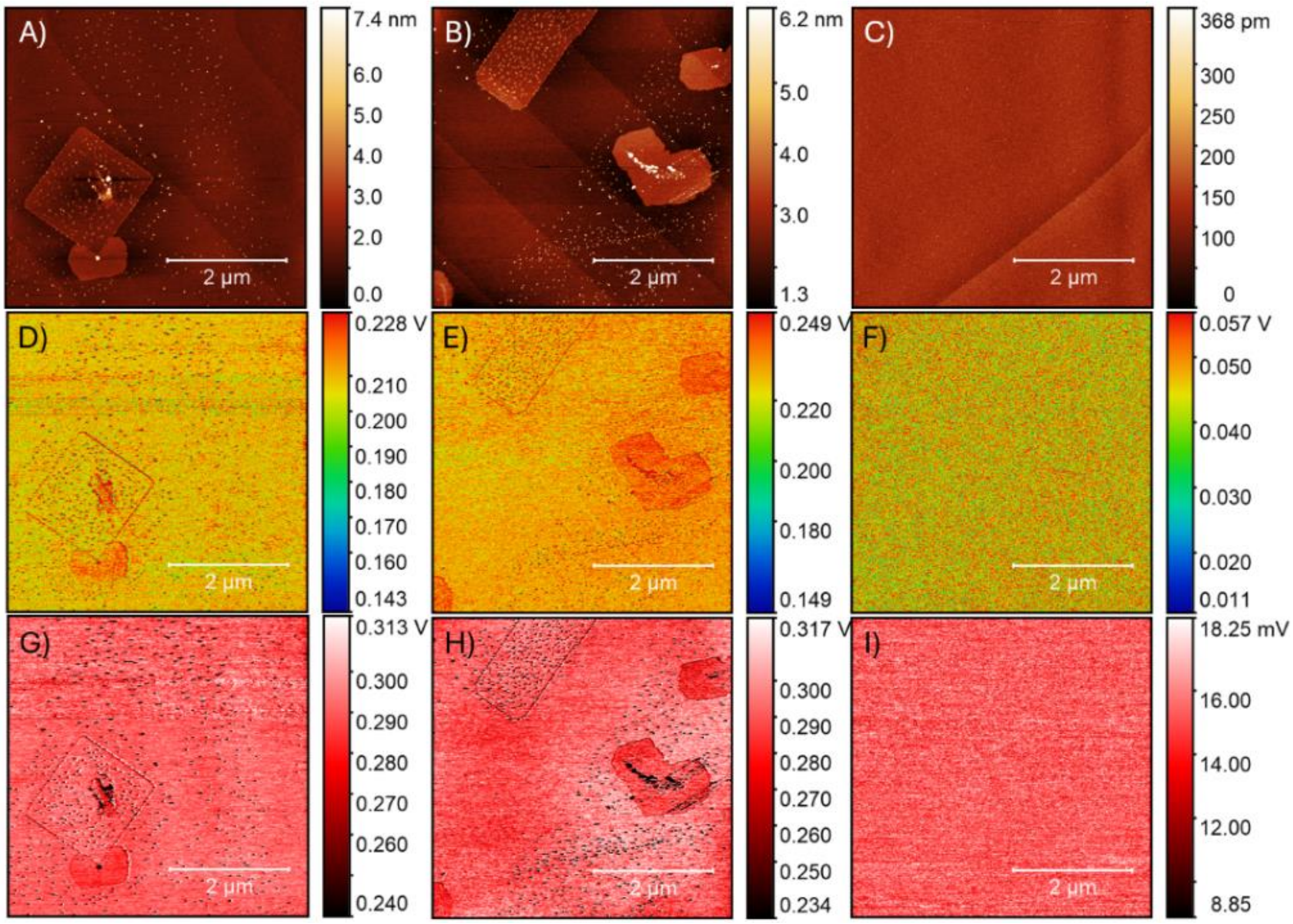

FIG 4. AFM PeakForce-QNM images of the freshly cleaved surfaces. Topography maps of sample A (Fgures A, B)) and sample B (Figure C) together with the corresponding nanomechanical maps showing the logarithm of the DMT modulus (Figures D–F) and adhesion to the AFM tip (Figures G–I). The small white points in Figures A and B are the protrusions described in paragraph c) above. For further details see Figure S11. The samples are air-stable for hours or days, but they change significantly after one year. Compare this to the AFM image of sample A after one year of storage in Figure S14 [24].

Although we cannot conclusively identify the origin of the defects, it is evident that the Se-poor, high-mobility sample is defect-free (see Figures 3C and 3F, Figures 4C and 4F, and Figure S12). The line defects in Figures 4B and 4C are attributable to dislocations or grain boundaries. These defects are abundant in the Se-rich sample but very rare in the Se-poor sample (see Figures 4B and 4C for comparison).

Preparing crystals with high mobility in the low-temperature region seems easy. We can rely on a selenium-poor ($V_{Se}$-rich) composition and the screening power of optical phonons. However, at room temperature, which is crucial for applications, electron-phonon scattering rather than point defects limits the relaxation time [12]. Now, let us analyze the features associated with highly mobile electrons at room temperature (Sample B). In this regard, $Bi_2O_2Se$ is a special case because mobile electrons are largely confined to Bi-derived states, i.e., the $Bi_2O_2$ layers in real space. It is important to note that the $Bi_2O_2$ layers only accommodate very high-frequency phonons (optical phonons only), while the Bi-Se bonding is mostly associated with low-frequency phonons. These two phonon groups have a large gap between them, so they have a very limited probability of interacting with each other [12].

The most important point is that the phonon-electron interaction decreases as the phonon frequency increases. The effective mass of the corresponding polaron remains close to the band mass [42]. Thus, the electron-phonon interaction is much weaker in the $Bi_2O_2$ layer than in the Se layer. This only applies if the Se-Bi derived phonons do not interact with the $Bi_2O_2$ derived phonons, i.e., if the Se-Bi derived phonons do not "penetrate" into the $Bi_2O_2$ layer. In perfect structure, this interaction is limited due to the large gap, as mentioned above. However, anharmonicities associated with point defects in the $Bi_2O_2$ layer can significantly increase this interaction because phonons only interact through anharmonicity. Therefore, we should observe a greater decrease in mobility with temperature in crystals with a defective $Bi_2O_2$ structure (see Figure 5b). Thus, room-temperature mobility is largely dictated by the perfection of the $Bi_2O_2$ layer rather than the Se layer. This model explains why the experimental room-temperature mobility (≈900 cm² V⁻¹ s⁻¹) can exceed the theoretical prediction (≈210 cm² V⁻¹ s⁻¹) [18]. This is due to the absence of low-frequency phonons in the perfect $Bi_2O_2$ layer. Unlike low-frequency phonons, high-frequency phonons can more effectively track electron motion. Therefore, we conclude that $Bi_{Se}$ defects limit carrier mobility in this peculiar manner at both low and room temperatures. The goal for room-temperature applications is therefore a crystal with a perfect $Bi_2O_2$ layer and a controllable $V_{Se}$ donor concentration. Section 9 of SM and ref. [43] provides examples of intentional doping into the Bi sublattice in single crystals, which can be compared with the undoped samples.

### E. Charge transport properties

We would like now to verify the coherence of the conclusions based on the structure and the band structure data with the transport and magneto-transport data. Figure 3 shows the electrical conductivity and Hall concentration and mobility of samples A and B. EDS analysis (see SM, section 2) indicates that both samples are slightly non-stoichiometric. Sample B is slightly Se-poor, and sample A is slightly Se-rich. This non-stoichiometry corresponds to a carrier concentration of the order of $10^{17}$ cm$^{-3}$. For example, an understoichiometry of $Se_{0.995}$ formally corresponds to ≈5 × $10^{19}$ cm$^{-3}$ if each defect produces one carrier. Therefore, the observed non-stoichiometry amply covers all the samples shown in Figure 1a.

Figure 5 shows that all parameters, electrical conductivity $\sigma$, Hall concentration $n_H$, and Hall mobility $\mu_H$ are lower for sample A than for sample B. The comparison of samples in terms of charge transport parameters including the SdH concentration $n_{SdH}$ (see Figure 6) is presented in Table III. Sample A clearly has worse charge transport and lacks traceable quantum (SdH) oscillations (see SM, section 4). These results are consistent with the above findings.

The $n_{SdH}$ concentration is higher than the $n_H$ concentration in sample B, which contradicts the findings in ref. [15] and the data for “relative” $Bi_2Se_3$ [44]. Due to the intrinsic anisotropy of this structure, this difference can be attributed to a non-spherical Fermi surface with respect to the band structure (see Figure 2). Unfortunately, the SdH effect cannot be analyzed for sample A (see Figure S4c and S4d in SM), which is due to low carrier mobility stemming from short relaxation times caused by defects. All samples analyzed in ref. [15] (Figure 1a) exhibited a pronounced SdH effect because their carrier concentration was higher than 5×$10^{17}$ cm$^{-3}$, i.e., they were in the high-mobility (Se-poor) region.

TABLE III. Comparison of transport parameters of samples A and B; electrical conductivity $\sigma$, Hall concentration $n_H$, SdH concentration $n_{SdH}$ and Hall mobility $\mu_H$. Sample C is included as an “intermediate” sample. This sample and the other samples are presented in SM [24], section 8, Figure S15.

| Temperature | ≈2 K | | | | ≈300K | | |
|---|---|---|---|---|---|---|---|
| Transport parameter | $\sigma$ ($\Omega^{-1}$m$^{-1}$) | $n_H$ (cm$^{-3}$) | $n_{SdH}$ (cm$^{-3}$) | $\mu_H$ (cm$^2$V$^{-1}$s$^{-1}$) | $\sigma$ ($\Omega^{-1}$m$^{-1}$) | $n_H$ (cm$^{-3}$) | $\mu_H$ (cm$^2$V$^{-1}$s$^{-1}$) |
| Sample A | 2.0 × $10^4$ | 5.0 × $10^{16}$ | no SdH | 2.4 × $10^4$ | 1.5 × $10^2$ | 5.0 × $10^{16}$ | 2.0 × $10^2$ |
| Sample B | 2.2 × $10^5$ | 1.5 × $10^{17}$ | 3.5 × $10^{17}$ | 9.4 × $10^4$ | 2.0 × $10^3$ | 1.4 × $10^{17}$ | 8.8 × $10^2$ |
| Sample C | 5.2 × $10^4$ | 9.3 × $10^{16}$ | - | 3.2 × $10^4$ | 5.7 × $10^2$ | 8.5 × $10^{16}$ | 4.2 × $10^2$ |

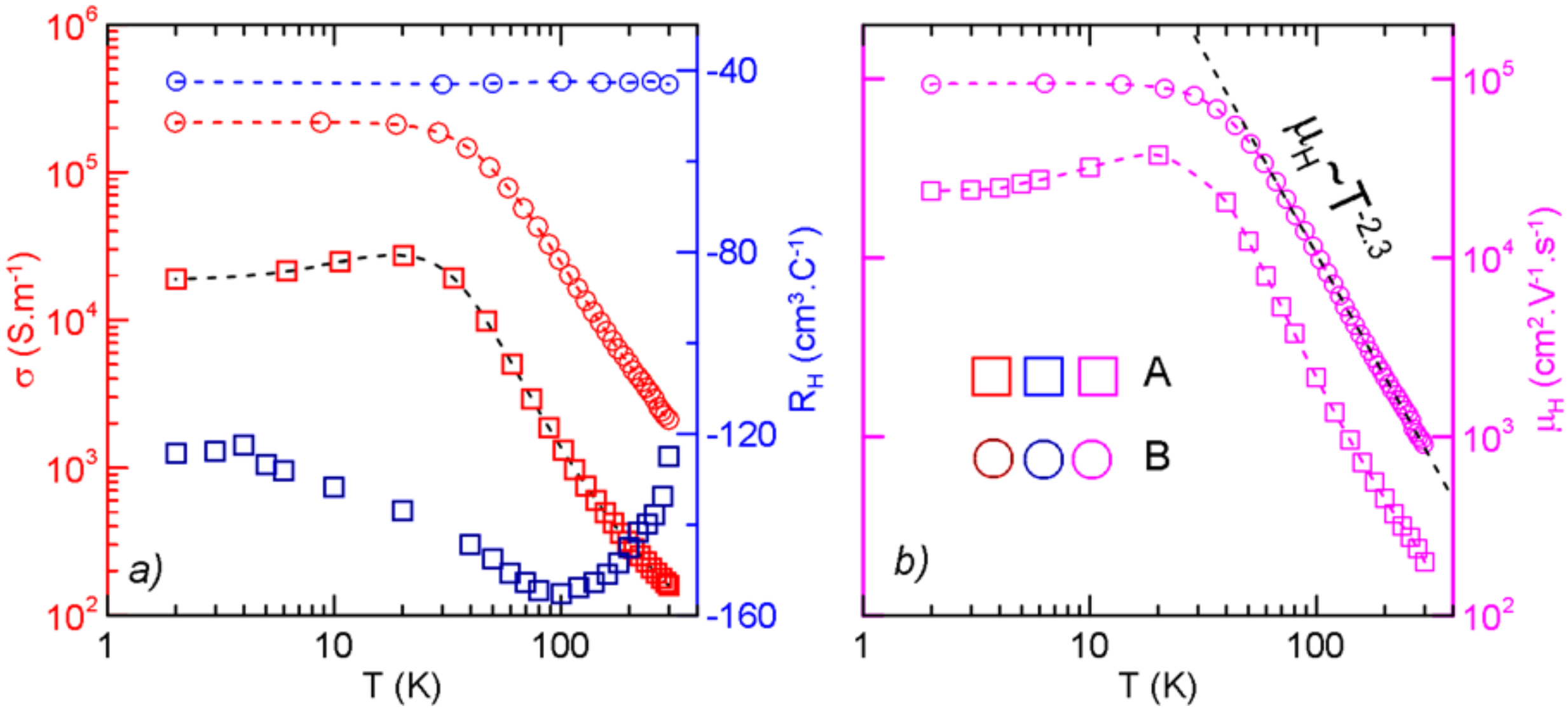

FIG 5. (a) Electrical conductivity $\sigma$ and Hall coefficient $R_H$ as a function of temperature between 2-300 K. (b) Their product, $R_H.\sigma$, gives the extremely high Hall mobility of electrons $\mu_H$, almost $\mathbf{10^5}$ $cm^2.V^{-1}.s^{-1}$ up to 30 K**.** High-temperature slope of the $\mu_H \approx T^{2.3}$ hints that the dominant scattering occurs on homopolar optical phonons which is intrinsic to layered materials [45,46]. We do not observe a T-square resistance, which likely only occurs due to higher electron concentrations of n>5 × $10^{17}$ $cm^{-3}$ [15].

The Hall coefficient, and thus the Hall concentration remains constant for sample B and depends very weakly on temperature for sample A. This is a typical metallic behavior, with no trace of activation up to 300K. This reflects the position of the active defects (e.g. $V_{Se}$, $Se_{Bi}$) with respect to the Fermi level and the large Bohr radius. Therefore, the electrical conductivity follows the temperature dependence of carrier mobility. The mobility remains very high (≈ 100 000 $cm^2V^{-1}s^{-1}$ in sample B) and nearly constant up to 30 K due to the screening of ionized impurities in a matter with high permittivity ($\varepsilon_r$≈500) [12]. In fact, one can neglect impurity scattering at any temperature with such a large permittivity [18]. It is worth noting that 30 K corresponds exactly to the temperature of phonon excitation in $Bi_2O_2$-layers [12]. This finding supports the concept of $Bi_2O_2$-based conduction channels and the discussion of high-temperature mobility above.

Sample A exhibits the room-temperature mobility $\mu_{HA}$≈ 200 $cm^2V^{-1}s^{-1}$, which is consistent with the predicted value of ≈ 210 $cm^2V^{-1}s^{-1}$ [18]. However, sample B has a very high mobility of ≈900 $cm^2V^{-1}s^{-1}$ and many other samples in Figure 1a have mobilities of up to ≈500 $cm^2V^{-1}s^{-1}$ at 300 K. At this temperature phonon scattering must dominate, and the ionized impurities must play a negligible role. Thus, the only direct explanation for the high mobility is a reduction of effective mass, $m^*$ and a less disordered structure as discussed above.

We aimed to address the possible presence of minority carriers in sample A in terms of mobility spectrum analysis. We measured Hall resistance and magnetoresistance for a set of temperatures up to 14 Teslas (see section 4 in SM). However, we left it open due to the possibility of overinterpretation. Figure 6 shows the resistive quantum oscillations extracted from the magnetoresistance measurement of sample B. The fast Fourier transform (FFT) of the data yields a single frequency of $f \cong 16$ T and a corresponding carrier concentration of $n_{SdH} \cong 3.5 \times 10^{17}$ $cm^{-3}$. From the analysis of the temperature-dependent data we obtain the cyclotron effective mass of $m_c^* \cong 0.13$, which is close to the estimation from THz cyclotron resonance of $m_c^* \cong 0.1$ (see SM [24], section 4). The Hall concentration $n_H \cong 1.5 \times 10^{17}$ cm$^{-3}$ is lower than the SdH concentration. In contrast, all samples analyzed in ref. [15] (Figure 1a) exhibit gradually lower SdH concentrations than the Hall concentration as the concentration decreases to ≈$6 \times 10^{17}$ cm$^{-3}$. This phenomenon is tentatively explained by the breakdown of Fermi sea into isolated lakes, contradicting the present result. The authors also report that T-square resistivity remains unexplained due to the absence of electron umklapp or inter-band scattering. Referring to Figure 2, we argue that a certain defect structure ($Se_{Bi}$) can induce multiple bands and, consequently, T-square resistivity. However, we do not observe this phenomenon significantly in our samples.

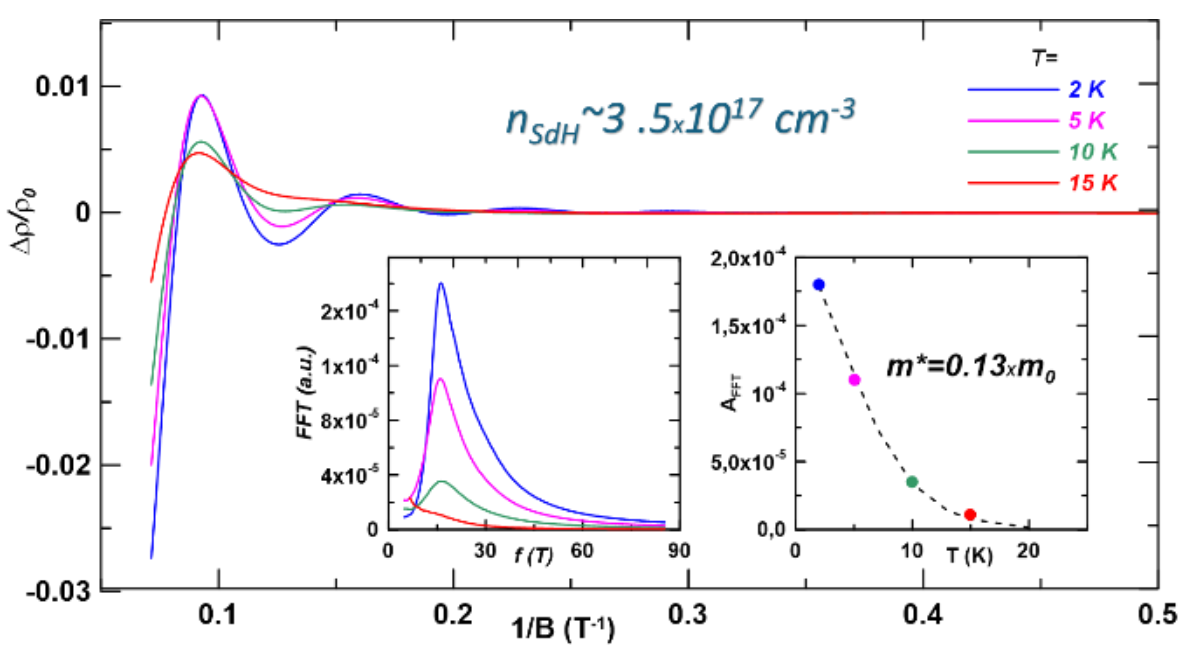

FIG 6. SdH curves revealing their single frequency at 15.6 T. The insets show the temperature dependent FFT amplitude, $A_{FFT}$ (left) and its dependence on temperature, T. The dashed line represents fit using the thermodynamic term of the Lifshitz-Kosevich equation and resulting cyclotron effective mass of $m_c^* \cong 0.13$. See SM [24], Figure S5, for rough data details.

## IV. CONCLUSIONS

The objective of this study was twofold: first, to explain the counterintuitive, step-like increase in Hall mobility with increasing carrier concentration; and second, to identify the factors that define high-temperature mobility.

We propose that the step-like increase primarily reflects a defect structure that alters the nature and effective masses of the carriers, consequently altering their mobility. Additionally, we demonstrate that Se-rich conditions are associated with multiple types of charge carriers, introducing ambiguity into Hall concentration and mobility. Indeed, Se-rich conditions can lead to a seemingly lower Hall concentration and mobility. This alone could explain the step-like increase in Hall mobility with concentration. Furthermore, Se-rich conditions are associated with the formation of a heterogeneous defect distribution, as well as macroscopic defects that compromise carrier mobility.

High-temperature mobility, which is governed by electron-phonon interactions, is primarily affected by $Se_{Bi}$ defects. These defects disrupt the periodicity and harmonicity of the $Bi_2O_2$ layer, which represents the primary charge conduction channel. Conversely, $V_{Se}$ and $O_{Se}$ do not destroy the $Bi_2O_2$ periodicity and harmonicity and they are associated with lower effective masses. Therefore, they do not deteriorate charge transport properties. Consequently, room-temperature carrier mobility is largely determined by the effective mass of carriers and the presence of minority carriers, if any. Both of these factors depend on the defect structure and the corresponding stoichiometry. Furthermore, the perfection of the $Bi_2O_2$ channel dictates not only the impurity scattering time, but also the electron-phonon scattering strength at elevated temperatures. Therefore, any doping in the form of substitution for Bi must harm electron mobility.

The goal for room-temperature applications is to create a crystal with a perfect $Bi_2O_2$ layer and controllable $V_{Se}$ donor concentration. To achieve this, the crystals must be grown at the optimal temperature and then cooled/annealed properly (see sample B). This raises the question of the optimal temperature. It must be high enough to maintain an adequate growth rate and $V_{Se}$ concentration, yet low enough to maintain a perfect $Bi_2O_2$ channel (800 – 830°C). Thus, the optimal temperature (and cooling rate) must be found to maintain a low $Bi_{Se}$ concentration and an adequate $V_{Se}$ concentration. Another important growth parameter is stoichiometry. Conditions with low Se content and high O content are recommended for high carrier mobility. The electron effective mass has been reported as low as $0.032m_0$ in Ref. [16]. This can serve as motivation.

## ACKNOWLEDGMENTS

This work was supported by the Czech Science Foundation (Project No. 22–05919S). This work was also supported by the Project TERAFIT - CZ.02.01.01/00/22_008/0004594 cofinanced by the European Union and the Ministry of Education, Youth, and Sports of the Czech Republic, and by the grant of the Ministry of Education, Youth, and Sports of the Czech Republic (Grant No. LM2023037).We are grateful to Vít Novák for supplying the GaAs sample. We gratefully acknowledge Dr. Hartmut Stadler (Bruker Nano Surfaces & Metrology, Karlsruhe, Germany) for performing the KPFM measurements.

# Optimizing high-temperature electron mobility in single-crystal $Bi_2O_2Se$ based on its unconventional dependence on concentration. Supplementary material

Antonín Sojka[1], Petr Knotek[1], Jan Zich[1], Martin Míšek[2], Roman Tesař[2], Kyo-Hoon Ahn[2], Petr Levinský[2] Jiří Navrátil[1], Pavlína Ruleová[1], Jiří Hejtmánek[2], Karel Knížek[2], Václav Holý[3,4], and Čestmír Drašar[1,*]

[1]*University of Pardubice, Faculty of Chemical Technology, Studentská 573,*

*53210 Pardubice, Czech Republic*

[2]*Institute of Physics of the Czech Academy of Sciences, Cukrovarnická 10/112,*

*162 00 Prague 6, Czech Republic*

[3]*Department of Condensed Matter Physics, Faculty of Mathematics and Physics, Charles University, Ke Karlovu 3, 121 16 Praha 2, Czech Republic*

[4]*Masaryk University, Department of Condensed Matter Physics, Kotlářská 2, 61137 Brno, Czech Republic*

[*]Corresponding author: cestmir.drasar@upce.cz

## 1 Hall effect and defects in $Bi_2O_2Se$ – unipolar vs. bipolar transport

Metallicity observed in $Bi_2O_2Se$ to very low carrier concentration can falsely indicate that we are dealing with a unipolar charge transport. We suggest that the unipolar transport surely applies to high enough concentration of electrons (n>$10^{17}$ $cm^{-3}$), which is associated with the Se-poor region as opposed to Se-rich region with the low carrier concentration [1]. For lower carrier concentrations, the nature and concentration of native defects are important. In particular, $Se_{Bi}$ creates defect bands within the gap, which eventually leads to semimetallic behavior, see Figure S9 ($Se_{Bi}$) and Figure S15. Note that the concentration of native defects in $Bi_2O_2Se$ can be high indeed; Bi/Se ratio = 1.4(se-poor) – 2.3 (Se-rich) as indicated by EDS analysis of structurally flawless samples in [2]. With respect to the high in-plane permittivity ($\varepsilon_r$ ≅500), the holes of any origin will form metallic itinerant system even for very low concentration of hole-like states [1].

Therefore, interpretation of the Hall effect in such samples can be complicated because both electrons and holes can contribute to the signal (Figure S4c). When interpreting Hall effect measurements, the same Hall coefficients can be obtained for very different carrier concentrations (Figure S1). This introduces uncertainty into the evaluation of Hall concentration and mobility for samples that may have an apparently high carrier concentration and therefore apparently low mobility.

Note that the opposite effect is also possible. In the case of two types of carriers of the same type. In particular, an "extra" increase of the Hall coefficient occurs when the ratio of the carrier concentrations is the inverse of the ratio of their mobilities (Figure S2). This is consistent with the band structure for the presence of $V_{Se}$ (Figure2 in the main text).  We do not consider Hall or anisotropy factors because they have a negligible effect on increasing carrier mobility with concentration.

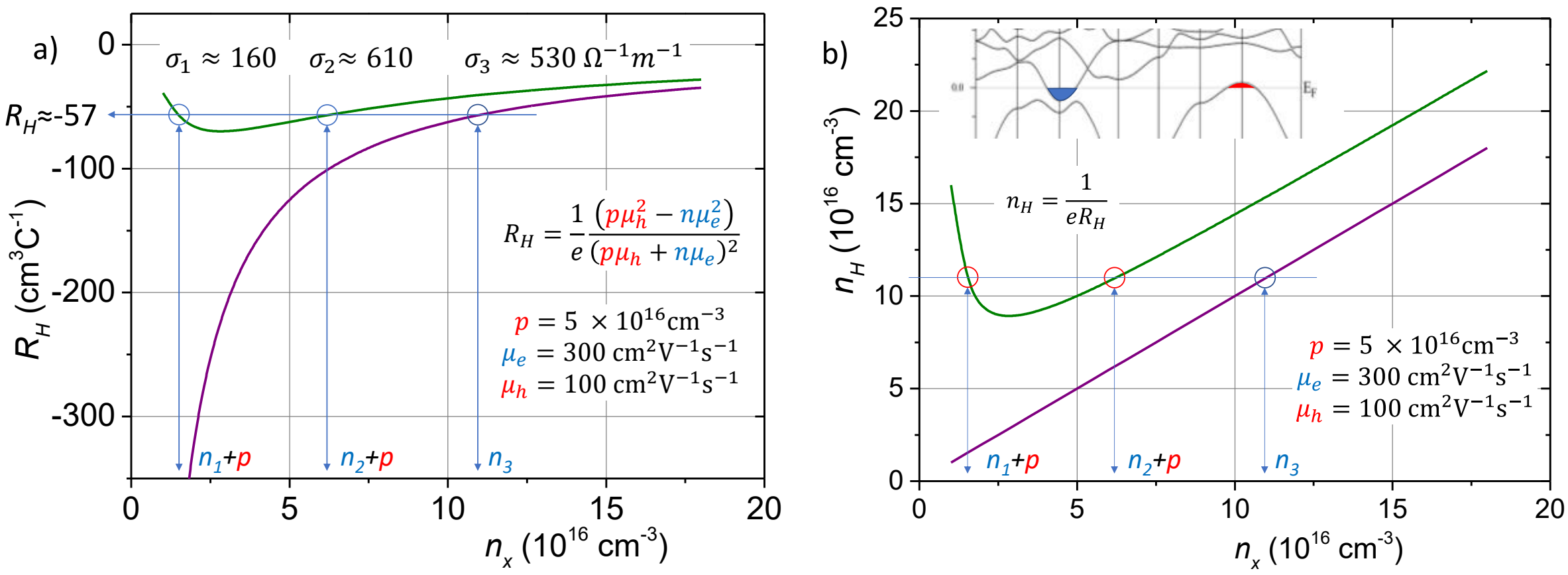

**Fig. S1**. a) Analysis of the Hall coefficient $R_H$ for one (violet curves) and two (green curves) types of charge carriers - electrons and holes. The hole concentration $p$ is fixed ($p$=5 x $10^{16}$ cm$^{-3}$). Three different charge carrier conditions ($n_3$≈11; $n_2$≈6, $p$=5 and $n_1$≈2, $p$=5 x $10^{16}$ cm$^{-3}$) can lead the same value of the Hall coefficient ($R_H$≈-57 cm$^3$C$^{-1}$) in this particular case, depending on the particular band structure ($n_3$≈unipolar while $n_2$+$p$ and $n_1$+$p$≈*bipolar).* b) These three conditions correspond to the same Hall concentration (11 x $10^{16}$ cm$^{-3}$) for unipolar transport. However, the correct Hall mobility $\mu_H = \sigma / e\, n_H$ can only be calculated for pure electron transport ($n_3$≈11 x $10^{16}$ cm$^{-3}$). For $n_2$≈6, $p$=5 and $n_1$≈2, $p$=5 x $10^{16}$ cm$^{-3}$ one obtains an unrealistically low electron mobility. Specifically, the corresponding increase in electrical conductivity $\sigma$ (in a) due to bipolar transport is too small to compensate for the artificially increased Hall concentration $n_H$ when calculating the Hall mobility. The band structure corresponds to the presence of $Se_{Bi}$ that can create a semimetallic band structure (GGA potential Figure S9). This corresponds to Se rich samples. The values of mobilities and concentrations are arbitrary.

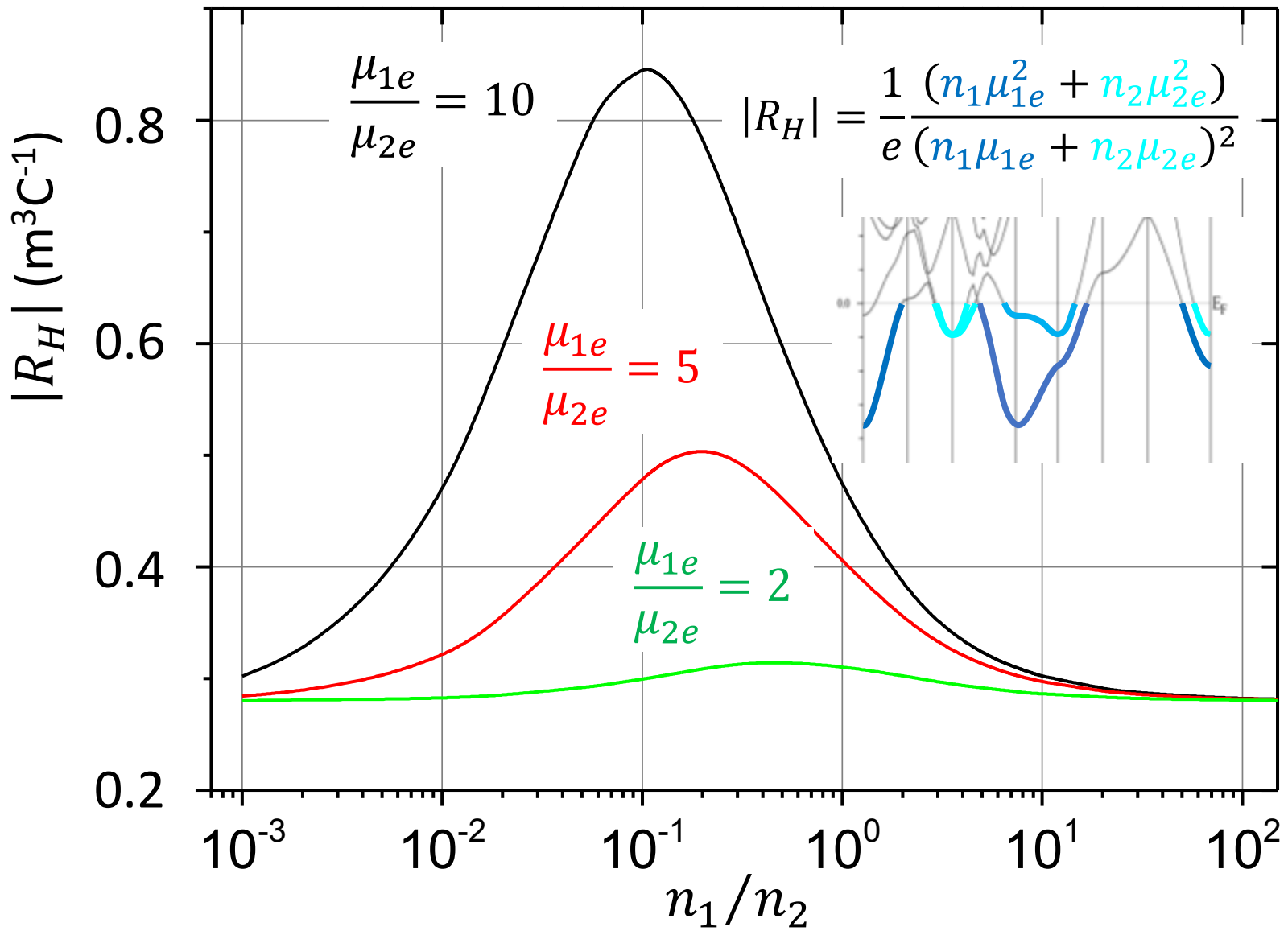

**Figure S2**. Analysis of the Hall coefficient $R_H$ for two types of electrons with different mobilities (masses). The presence of two types of electrons can lead to a significant increase in RH and correspondingly to apparently lower Hall concentration. The maximum increase occurs when the inverse of their concentration ratio is their mobility ratio. This can lead to an overestimation of electron mobility. The band structure corresponds to the presence of $V_{Se}$ that is associated with the occurrence of two (or more) types of electrons. This corresponds to Se poor samples (GGA potential, Figure S9). The values of mobilities and concentrations are arbitrary.

## 2 EDS elemental analysis

Energy dispersive analyses were performed on the samples to determine the true stoichiometry of the crystals (in the form of $Bi_{2\pm x}O_{2\pm y}Se_{1\pm z}$) and compare them with the conclusions drawn from HRXRD. The averaged results (six spot measurements distributed over the sample area) give the stoichiometry $Bi_{1.997}O_{1.998}$ $Se_{1.005}$ (±0.01, 0.02, 0.01, respectively) for Se-rich and $Bi_{1.991}O_{2.014}$ $Se_{0.995}$ (±0.04, 0.06, 0.02) for Se-poor. We see that the non-stoichiometry is much smaller than predicted by HRXRD (see section 3 below). On the other hand, variations in the Bi/Se ratio of 1.4 to 2.3 have been reported in the literature [2] for apparently perfect crystals with a very high carrier concentration ($>10^{20}cm^{-3}$). This suggests that a small shift in the stoichiometry ratios can induce changes in the defect chemistry with a significant effect on charge transport. Although it seems unlikely, these findings suggest that combinations of defects capable of preserving the original stoichiometry exist, for example $3V_{Se} + 2Se_{Bi} + 2V_O$ or $3O_{Se} + 2Se_{Bi} + 5V_O$. The presence or absence of $Se_{Bi}$ defects can be crucial in terms of charge transport, as discussed in the main text, section B, as they eventually induce the semimetal state and thus bipolar transport. However, $V_{Se}$ is the main donor below 300 K and its concentration should be comparable with electron concentration [3], which is of the order of $10^{17}$ $cm^{-3}$, according to the Hall measurement (see main text). This corresponds for example to z ranging from 0.00001 to 0.0001 in the $Bi_2O_2Se_{1-z}$ notation, an immeasurably small concentration.

Table S2A: EDS analysis of Se-rich sample used for characterization in the main text (Sample A).

| Measurement no. | Bi (at.%) | O (at.%) | Se (at.%) | Corresponding stoichiometry | Bi | O | Se |
|---|---|---|---|---|---|---|---|
| 1 | 39.32 | 20.32 | 40.36 | | 2.02 | 1.97 | 1.02 |
| 2 | 40.97 | 19.71 | 39.32 | | 1.97 | 2.05 | 0.99 |
| 3 | 39.58 | 20.21 | 40.21 | | 2.01 | 1.98 | 1.01 |
| 4 | 39.95 | 20.14 | 39.92 | | 2.00 | 2.00 | 1.01 |
| 5 | 39.67 | 20.22 | 40.1 | | 2.01 | 1.98 | 1.01 |
| 6 | 40.25 | 20 | 39.75 | | 1.99 | 2.01 | 1.00 |
| | | | | Average | **1.997** | **1.998** | **1.005** |
| | | | | Error ± | 0.01 | 0.02 | 0.01 |

Table S2B: EDS analysis of Se-poor sample used for characterization in the main text (Sample B).

| Measurement no. | Bi (at.%) | O (at.%) | Se (at.%) | Corresponding stoichiometry | Bi | O | Se |
|---|---|---|---|---|---|---|---|
| 1 | 41.34 | 19.5 | 39.16 | | 1.96 | 2.07 | 0.98 |
| 2 | 38.92 | 20.42 | 40.66 | | 2.03 | 1.95 | 1.02 |
| 3 | 38.66 | 20.41 | 40.94 | | 2.05 | 1.93 | 1.02 |
| 4 | 40.49 | 19.92 | 39.59 | | 1.98 | 2.02 | 1.00 |
| 5 | 42.03 | 19.24 | 38.74 | | 1.94 | 2.10 | 0.96 |
| 6 | 41.34 | 19.5 | 39.16 | | 1.99 | 2.01 | 0.99 |
| | | | | Average | **1.991** | **2.014** | **0.995** |
| | | | | Error ± | 0.04 | 0.06 | 0.02 |

## 3 High resolution X-ray diffraction

We conducted high-resolution X-ray diffraction (HRXRD) experiments to compare the structural quality of the two single crystals under investigation. Specifically, we compared the angular mosaicity, domain thickness and the overall sample quality. Figure S3a shows long symmetric $2\Theta/\omega$ scans of both samples, along with the results of the fits of the heights of the diffraction maxima.

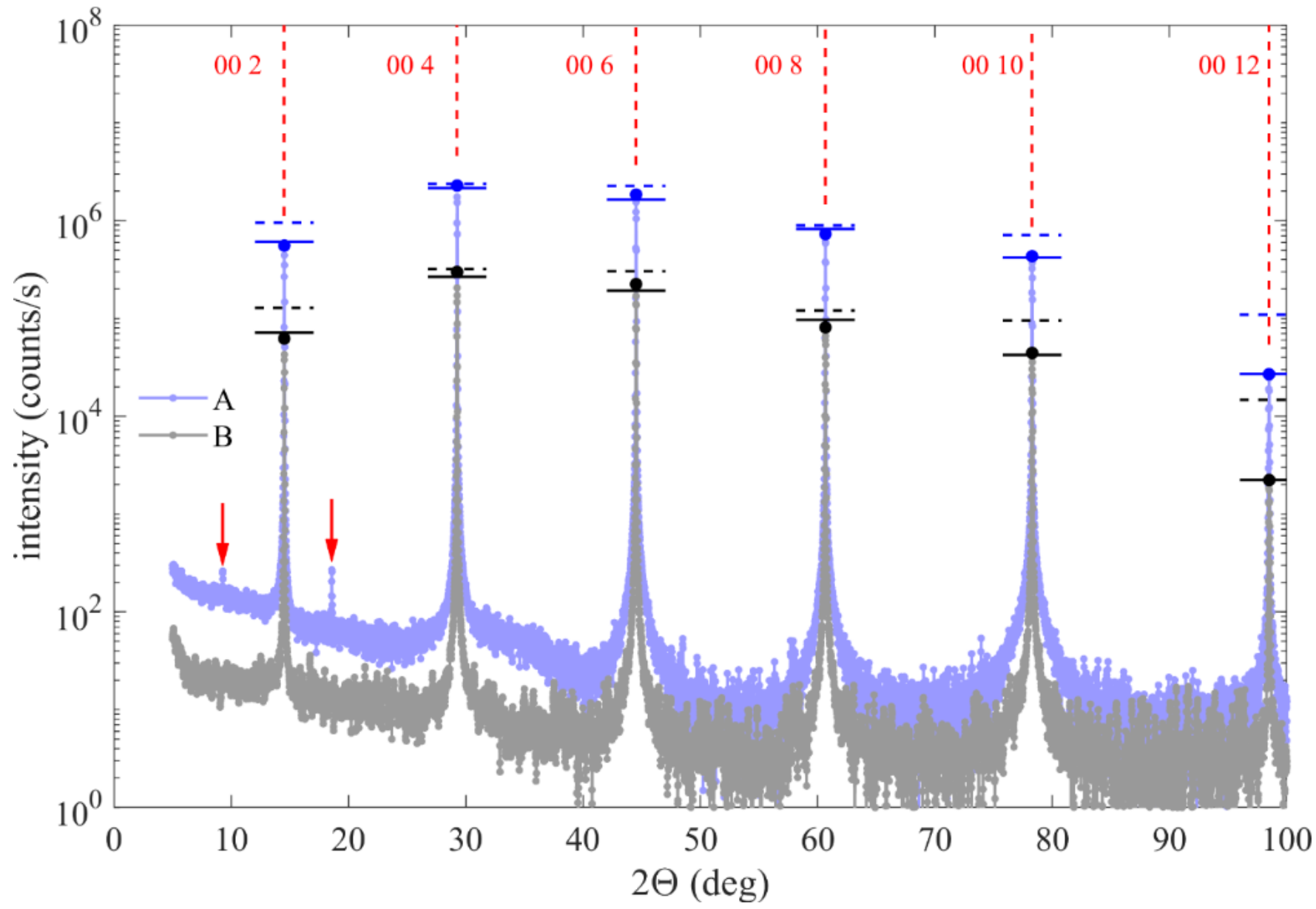

**Figure S3a**. HRXRD symmetric 2Θ/$\omega$ scans of the samples A and B. The blue and black large dots emphasize the measured intensity maxima. The short horizontal dashed lines represent the maxima following from the nominal tetragonal $Bi_2O_2Se$ lattice [4]. The horizontal solid lines indicate the fitted maxima. The red vertical dashed lines indicate the theoretical diffraction maxima 00(2n). The two samples are almost identical. Two very weak reflections in the sample A curve, 2Θ≈9° and 18° can be tentatively attributed to monoclinic Se (2Θ ≈ 9°) and $Bi_2SeO_5$ (2Θ ≈ 18°).

We fitted the heights of the diffraction maxima to a structural model to determine the fractional excess selenium (Se) density in the bismuth (Bi) lattice, $x_{Se}$, and the distance between the Bi and Se (001) planes.

In detail, the maximum intensity is fitted; this is calculated within the framework of modified kinematic theory, absorption into account. A modified Lorentz factor, L = 1/sin(theta)/sin(2*theta), was used, which is suitable for mosaic crystals with significant disorientation of the mosaic blocks.

The structural model included the following fitted parameters:

1. The distance between the Se and $Bi_2O_2$ planes
2. Thermal B factor
3. occupancy of O sites
4. occupancy of Bi sites
5. Se site occupancy
6. The relative amount of Se at Bi sites.
7. The relative amount of Bi at Se sites.
8. Primary intensity
9. Background

The lattice parameters were determined directly from the diffraction angles and entered into the fit as fixed parameters. As the amount of input data (diffraction peak intensities) was relatively small, the free parameters were correlated and could not be fitted simultaneously. Ultimately, we only fitted parameters 1, 2, 6, 8 and 9. An equally good fit would have been obtained if parameters 4 and 5 had been fitted instead of parameter 6. This is because the DFT calculation and literature data suggest that $Se_{Bi}$ and $V_{Se}$ are defects with the lowest formation energy.

The resulting parameters were nearly identical for both samples: z_(Bi-Se) = (0.355 ± 0.001)c (the nominal value is 0.3534 for both samples),and $x_{Se} = 0.056 \pm 0.033$ ($Bi_{1.89}O_2Se_{1.11}$) for samples A and $x_{Se} = 0.049 \pm 0.024$ ($Bi_{1.90}O_2Se_{1.10}$) for samples B. It is important to note that a slight decrease in the scattering power of the Bi atoms (modeled by replacing of Bi with lighter Se atoms) can have various causes, such as Bi vacancies, $V_{Bi}$ or O atoms on Bi sites, $O_{Bi}$. Unfortunately, the measured data cannot distinguish between these scenarios. However, given the formation energies of defects, this scenario is the most likely.

We used the positions of the diffraction maxima to determine the vertical lattice parameter c exactly, using the Cohen-Wagner method [5], and we estimated the vertical thickness L of coherent domains using the Williamson-Hall method [6]. The measured plots together with the determined parameters c and L are shown in Figure S3b.

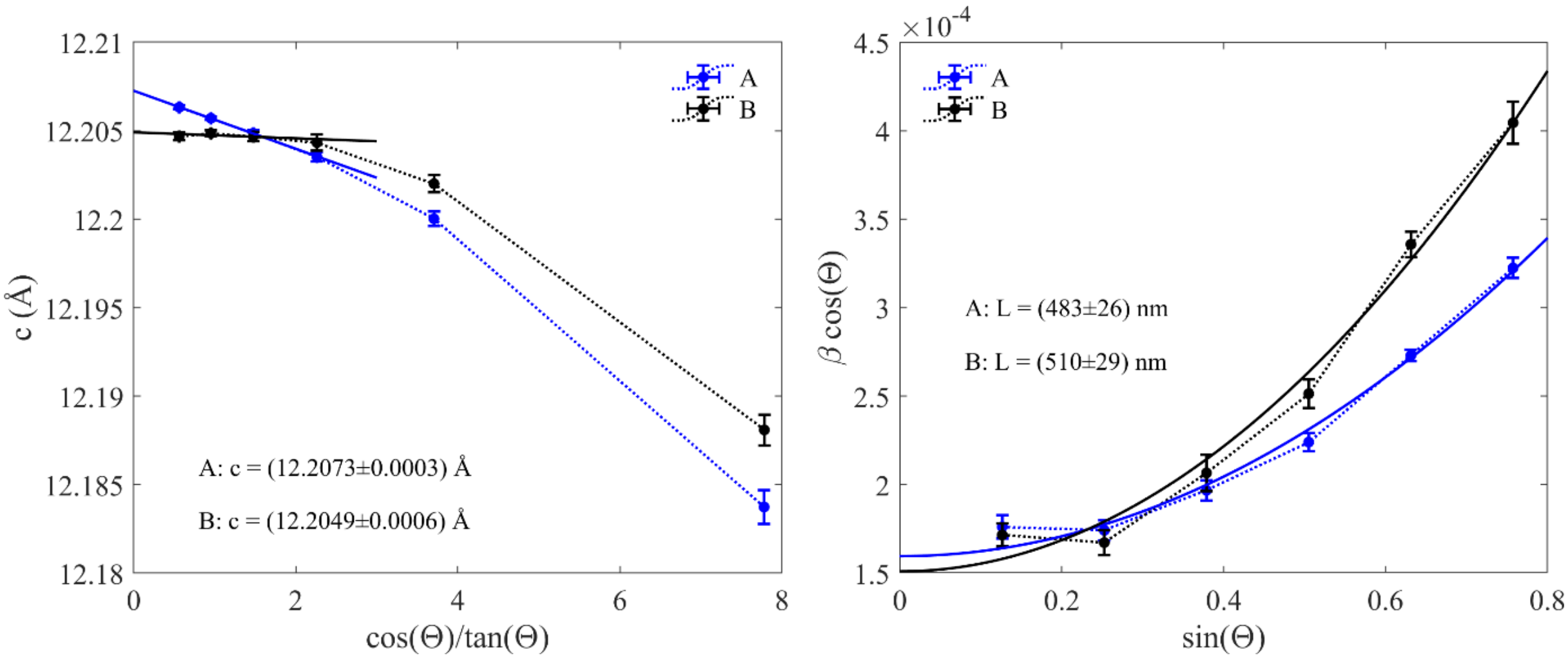

**Figure S3b**. The Cohen-Wagner plot (left) and the Williamson-Hall plot (right) of samples A and B, along with the values of vertical lattice parameter c and the vertical size L of the coherent domains.

Indeed, the structures of both samples are very similar. The vertical lattice parameter of sample A is slightly larger by 0.002 Å, which is only slightly above the margin of error but could indicate the presence of embedded extraneous phases (see Figure 3 in the main text). The shapes of the Williamson-Hall plots are unusual. Typically, the dependence of $\beta \cos \Theta$ on $\sin \Theta$ is linear. The reason for the obvious nonlinearity is unclear. Nevertheless, the vertical size L of the domains is calculated by extrapolating the graph to zero abscissa. The value for sample B is slightly larger than that for sample A; however, the difference lies within the error margin. A domain size of 0.5 μm is comparable to the coherence width of the primary X-ray beam; therefore, this value represents rather the lower limit of the true domain size.

Figure S3c shows the reciprocal-space maps of both samples, which were measured around the reciprocal lattice point (0 0 6), with the intensities integrated along the radial 2Θ/ω axis. We can deduce the angular misorientation of the mosaic blocks from the integrated intensities. Sample A has a broader main maximum by about 0.2°, whereas sample B exhibits a long, weak streak in the ω-direction, indicating the presence of a few very small, highly misoriented blocks.

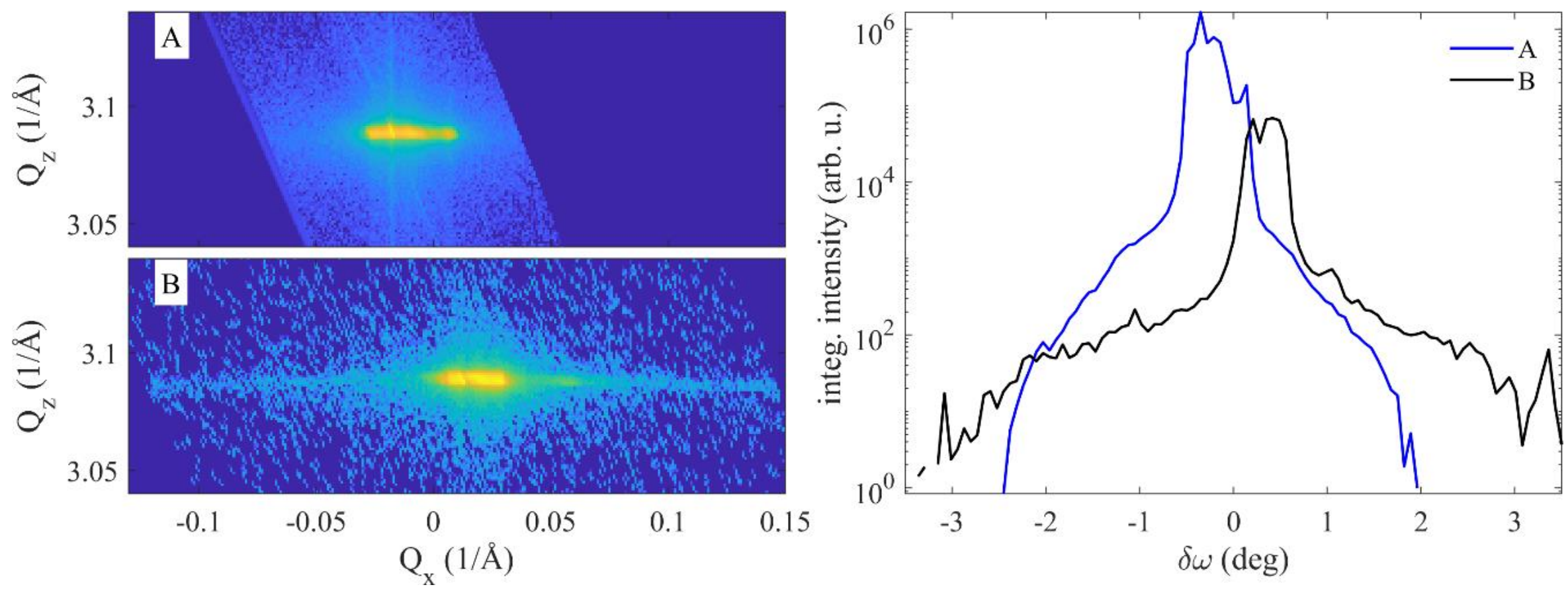

**Figure S3c**. Reciprocal-space maps around 0 0 6 (left) and the intensity integrated along the radial $2\Theta/\omega$ axis (right). The colors in the reciprocal-space maps span over 5 decades.

Finally, Figure S3d shows the asymmetric reciprocal-space maps taken around the reciprocal lattice point 2 0 10. We integrated the measured maps along the Debye rings ($2\Theta = const$; the rings crossing the diffraction maxima are shown by red lines). From the positions of the maximum of the integrated intensity on the $2\Theta$ axis as well as from the parameters c determined above, we calculated the in-plane lattice parameter a.

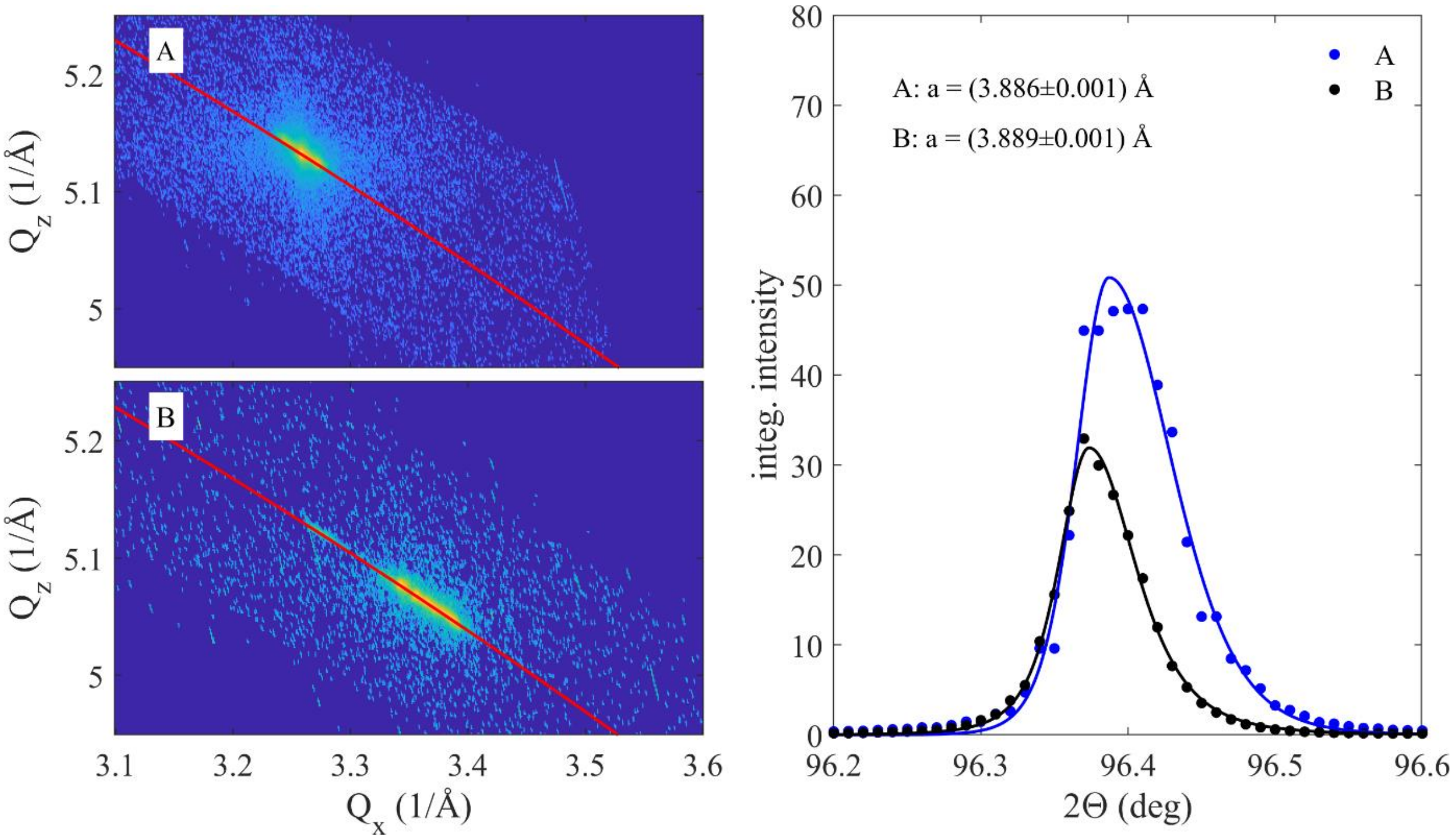

**Figure S3d**. Asymmetric reciprocal-space maps taken around 2 0 10 reciprocal lattice point (left) and their intensities integrated along the Debye circles $2\Theta = const$ (right). The red lines in the left panels represent the Debye circle crossing the diffraction maximum, the colors in the maps span over 5 decades.

The results shown in Figures S3 a-d suggest that the main difference between the two samples appears to be the crystal quality, which is associated with the defect structure. The difference is minimal, with sample A being slightly more tetragonal. This hypothesis is supported by variations in carrier mobility, electrical conductivity (see main text) and cyclotron resonance (Figure S4). It is also supported by DFT calculations using the GGA potential, albeit only weakly. Specifically, relaxing the atomic positions reveals a greater lattice parameter a for the $V_{Se}$ defect (a=0.3907 nm) than for the $Se_{Bi}$ defects (a=0.3833 nm). However, there is a discrepancy in the carrier concentration, which is counterintuitively lower in the more defective structure of sample A. This discrepancy is one of the problems solved in this work.

## 4 Cyclotron resonance measurement in FIR/THz region, magnetotransport

Since the magnetoresistance yielded different results for samples A and B (see main text), we performed cyclotron resonance experiments with circularly polarized THz radiation. The magnetic field was applied perpendicular to the sample plane and parallel to the incident laser beam. We measured the transmission and reflection of the samples while varying the magnetic field. As the cyclotron frequency is proportional to the applied field, the resonance condition can be achieved at a fixed laser frequency.

However, the observation of a resonant state depends on several additional factors, including the direction of circular polarization, the polarity of the magnetic field, the type and lifetime of the charge carriers, as well as the sample temperature. We tested the experimental configuration by probing the cyclotron resonance in a reference sample, a two-dimensional electron gas (2DEG) confined in a GaAs quantum well. As shown in Figure S4a, the measured transmission exhibits a clear resonance signal in a positive magnetic field but remains flat in a negative field, which confirms the high purity of the circular polarization. For the reversed sense of the circular polarization, the resonance emerges in a reversed magnetic field (see the inset in Figure S4c). In the case of hole-type charge carriers, this picture is exactly the opposite.

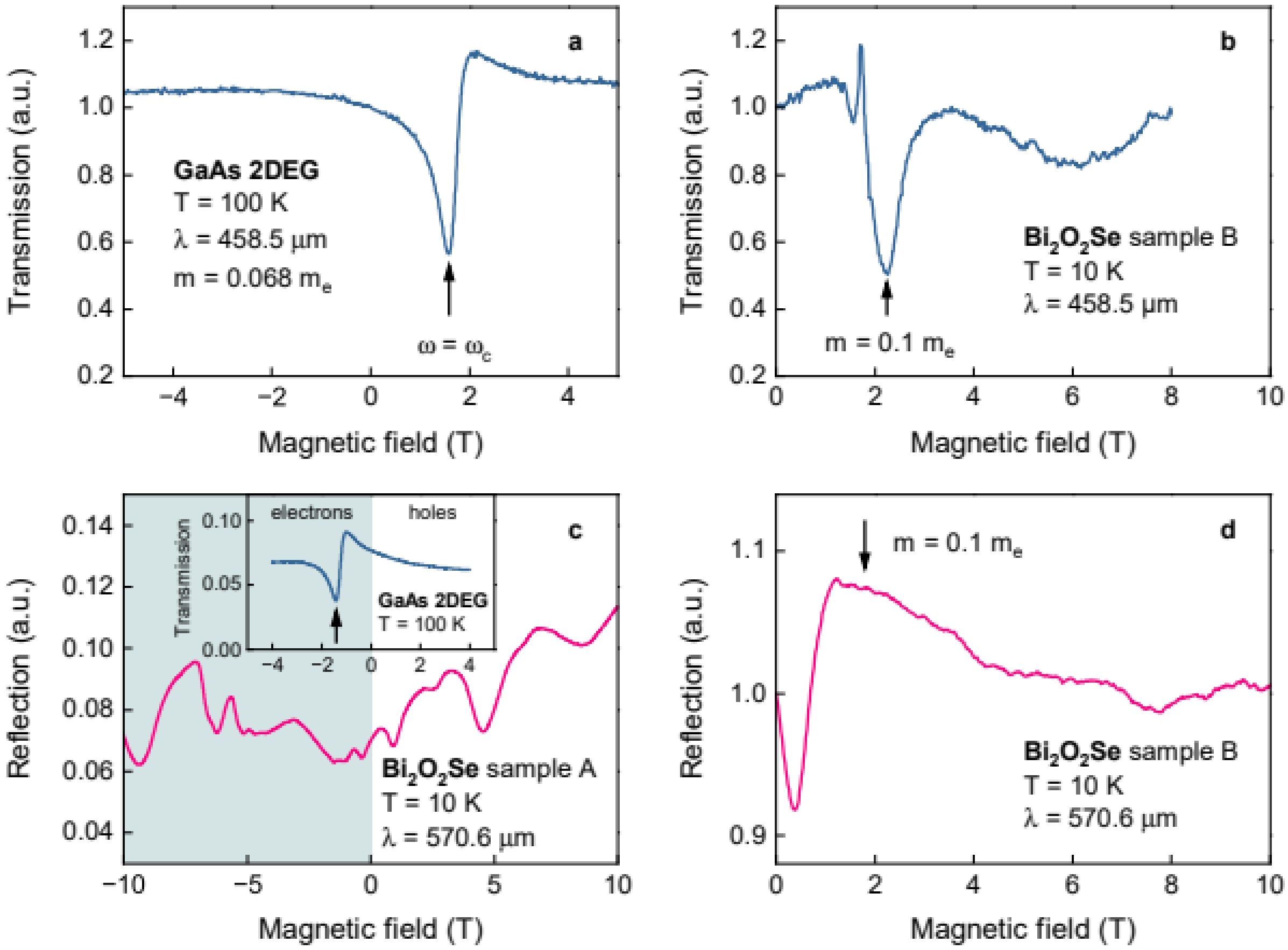

**Figures S4** (a, b) Cyclotron resonance in a GaAs-2DEG reference sample and $Bi_2O_2Se$ sample B observed in transmission mode. (c,d) Cyclotron resonances in $Bi_2O_2Se$ sample A and sample B in reflection mode. The distinct cyclotron resonance corresponding to an effective electron mass of about 0.1 $m_e$ is observed in sample B, whereas many signals are observed for the sample A. For sample B, the reflectivity in the active branch is plotted divided by the reflectivity in the inactive branch, i.e., R(B)/R(-B). This eliminates unwanted signal hysteresis caused by interference between the sample and the cryostat windows.

Cyclotron resonance measurements of $Bi_2O_2Se$ samples A and B are displayed in Figures S4b-d. Sample A shows a rather complex pattern with many minima and maxima, likely due to a combination of interference fringes and cyclotron resonance on various charge carriers. The increased interference effects can be explained by the lower electrical conductivity, which results in less attenuation of the THz rays that are reflected multiple times within the sample plate. The reference measurement, shown in the inset, defines the regions of electron and hole resonances. These preliminary results suggest the presence of different types of carriers in sample A. However, further experiments with additional laser lines are necessary to confirm this finding.

Transmissions of the GaAs-2DEG reference and $Bi_2O_2Se$ sample B were obtained under similar experimental conditions (see Figures S4a and S4b). The resonances occur at the minimum transmission, as indicated by the arrows. The electrical conductivity of the $Bi_2O_2Se$ sample B can be attributed to electron charge carriers with an effective mass of about $m_c^*$≈0.1. Cyclotron resonance is also visible in the reflection measurements, as shown in Figure S4d. In contrast to transmission, the resonance occurs near the maximum reflection, as marked by the arrow position for the given effective mass and laser line. It should be noted that the optical properties can be affected by the multiple reflections within the sample. In this case, the minimum of transmission and the maximum of reflection do not correspond exactly to the resonance condition, but show a slight offset, which is more significant for the reflection.

Figure S5a shows a comparison of rough magnetoresistance data for samples A and B at a temperature of 2 K. Unlike sample B, sample A shows no traceable oscillations.

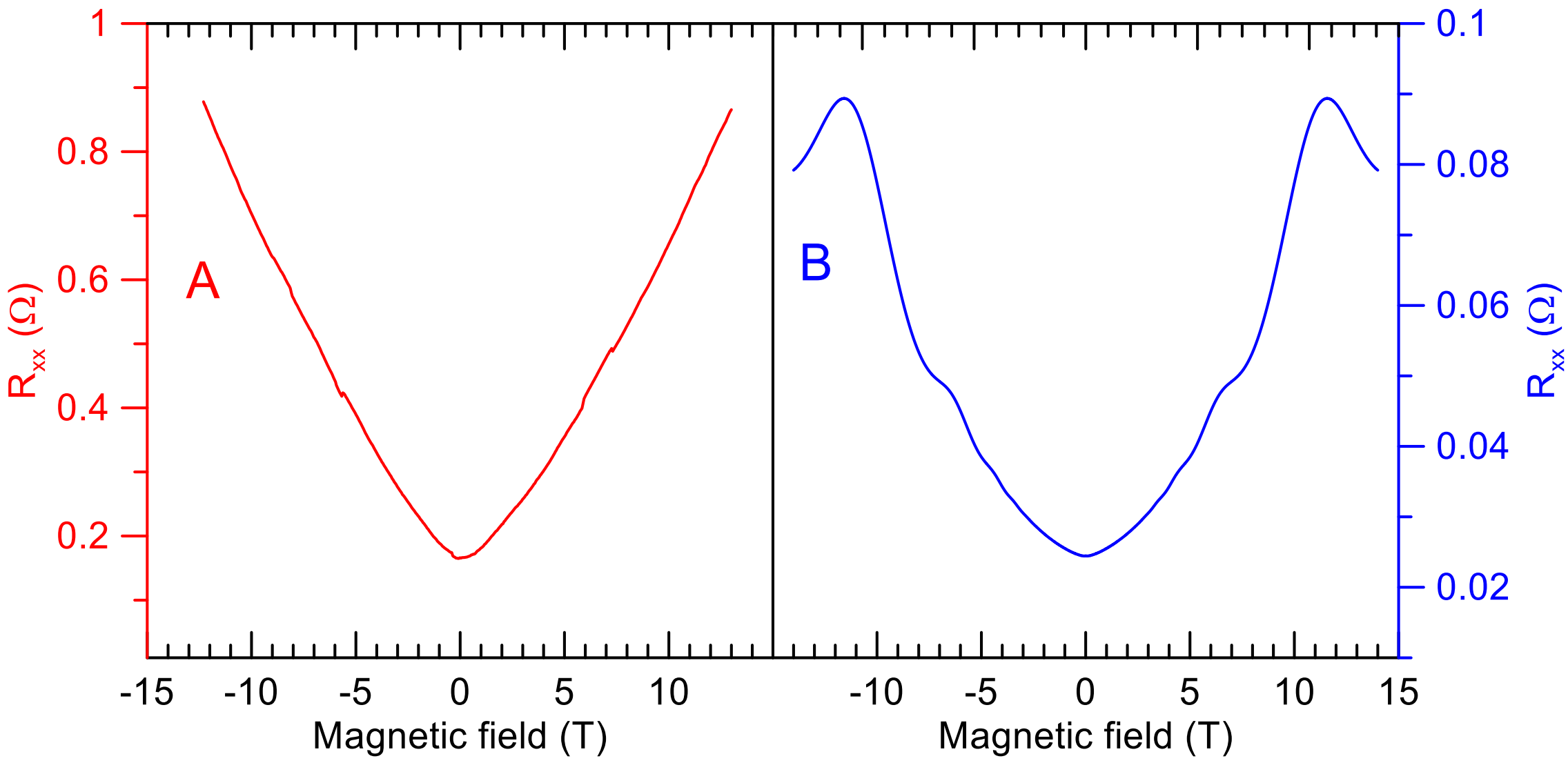

**Figure S5a** Comparison of rough magnetoresistance data for samples A and B.

We measured the Hall resistance and magnetoresistance for a set of temperatures up to 14 Teslas (Figure S5b) to address the possible presence of minority carriers in sample A (see Figure S4c) in terms of mobility spectrum analysis.

The linearity and minimal evolution between 2 and 30 K do not allow for a "doubtless" serious analysis without overinterpretation. However, since the Hall data differs only slightly when the temperature changes by more than 10x, we can assume that the carriers are nearly temperature-independent, assuming a multicarrier system (e.g., deduced from the complex THz spectra in Fig. S4c). Otherwise, the evolution of their concentration/mobility would be "exotically" compensated. This picture aligns with theoretical calculations of the density of states (DOS) for potential defects in Se-over-stoichiometric $Bi_2O_2Se$, which suggest the presence of "midgap" states (see Figures S6 and S7). We provide the data for interested readers.

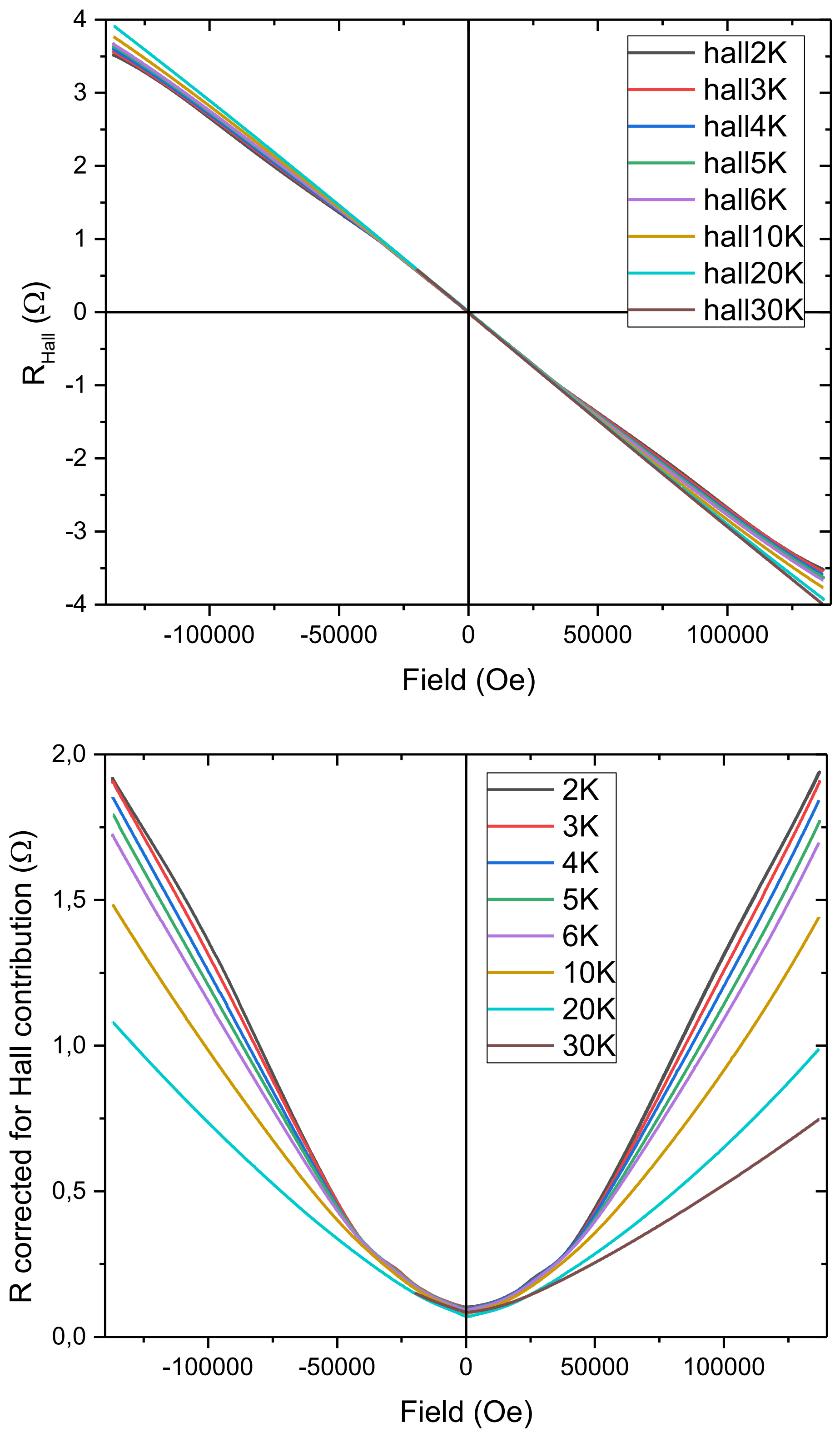

**Figure S5b** Hall resistance and magnetoresistance of sample A at various temperatures. Raw data was measured on Sample A with the following dimensions: length x width x thickness = 2.29 x 2.1 x 0.48 mm$^3$.

## 5 Energy formation of native defects

Table S5 summarizes formation energies of native defects in $Bi_2O_2Se$. On average, the $O_{Se}$, $V_{Se}$ and $Se_{Bi}$ have reasonably low formation energies. Therefore, we consider these three defects and neglect any occurrence of other native defects.

Table S5: Formation energies of native defects in $Bi_2O_2Se$, calculated using two potentials: mBJ and GGA.

| Description | Stoichiometry of 3×3×1 cell | mBJ | GGA |
|---|---|---|---|
| Stoichiometric | $Bi_{36}O_{36}Se_{18}$ | 0.00 | 0.00 |
| Substitution O for Se, $O_{Se}$ | $Bi_{36}O_{37}Se_{17}$ | 0.99 | -0.05 |
| Vacancy Se, $V_{Se}$ | $Bi_{36}O_{36}Se_{17}$ | 2.65 | 4.77 |
| Substitution Se for Bi, $Se_{Bi}$ | $Bi_{35}O_{36}Se_{19}$ | 3.27 | 3.76 |
| Substitution Se for O, $Se_O$ | $Bi_{36}O_{35}Se_{19}$ | 4.29 | 5.25 |
| Vacancy O, $V_O$ | $Bi_{36}O_{35}Se_{18}$ | 4.40 | 6.55 |
| Vacancy Bi, $V_{Bi}$ | $Bi_{35}O_{36}Se_{18}$ | 8.93 | 6.67 |

## 6 Changes of band structure due to native defects

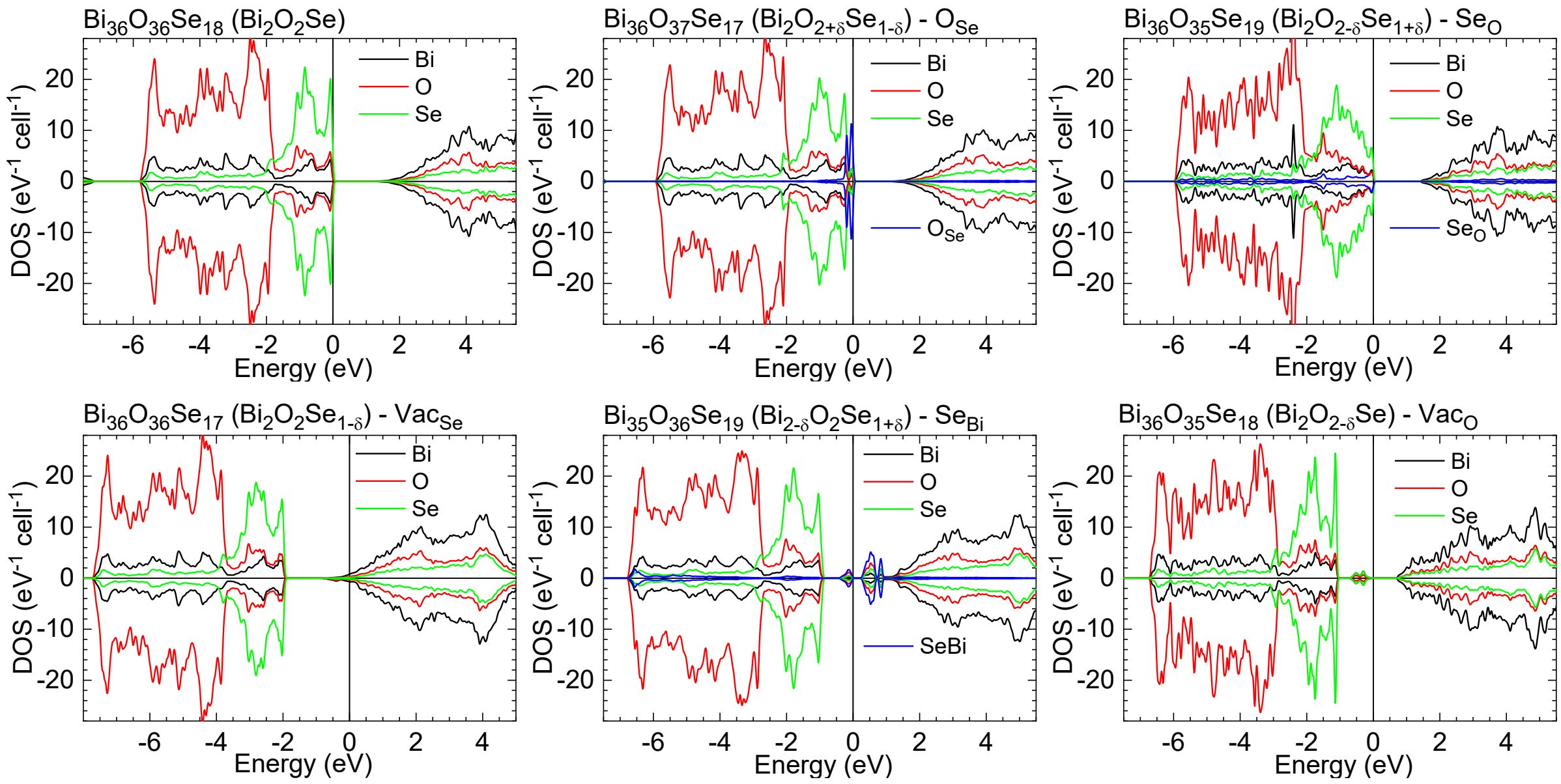

**Figure S6**. Calculated DOS for stoichiometric $Bi_2O_2Se$ and five native defects with the lowest formation energy, using mBJ potential. Three of these defects show significant effect on the band structure. $Se_{Bi}$ and $V_O$ form states within the bandgap and $V_{Se}$ shifts the Fermi level into conduction band.

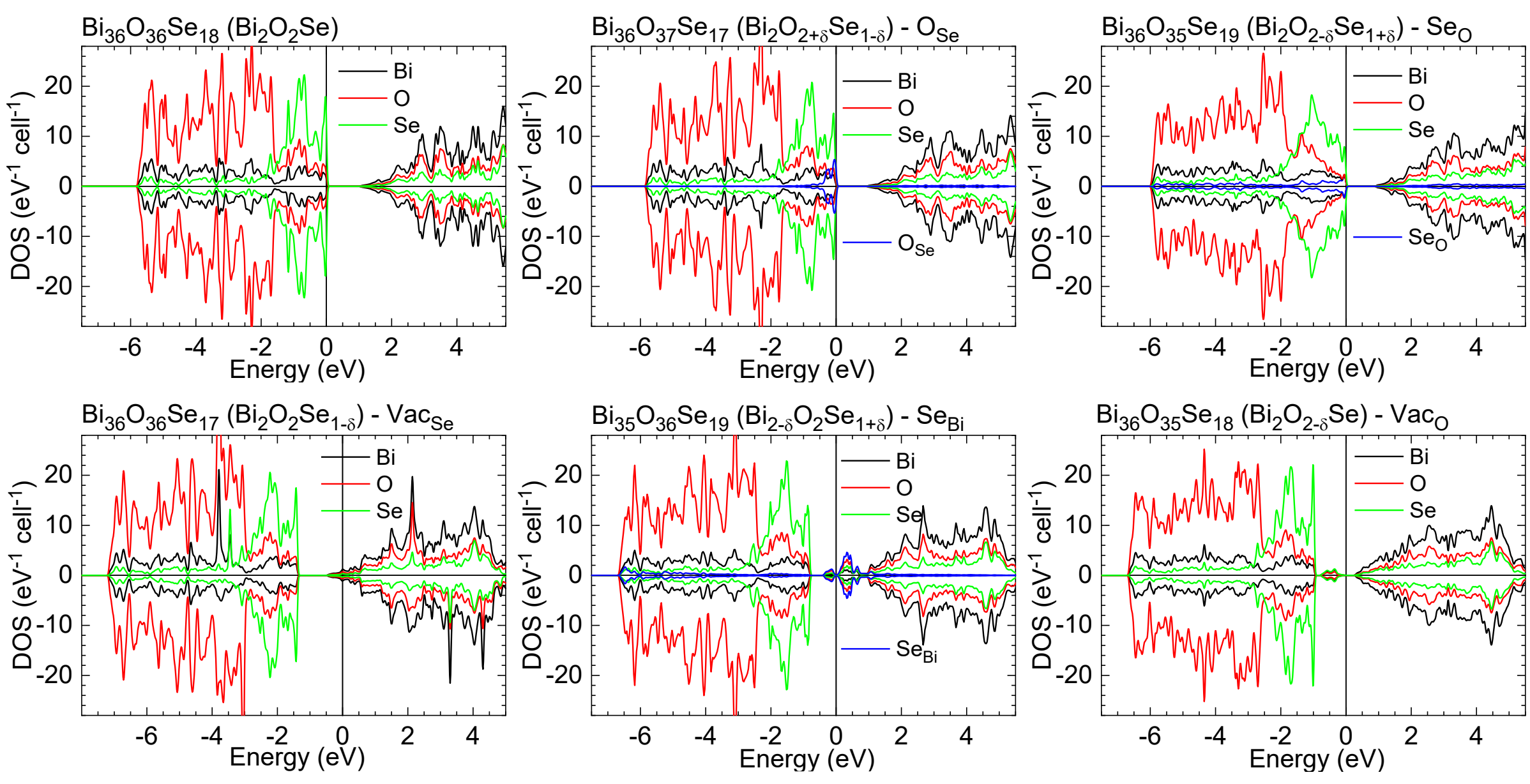

**Figure S7**. Calculated DOS for stoichiometric $Bi_2O_2Se$ and five native defects with the lowest formation energy, using GGA potential.

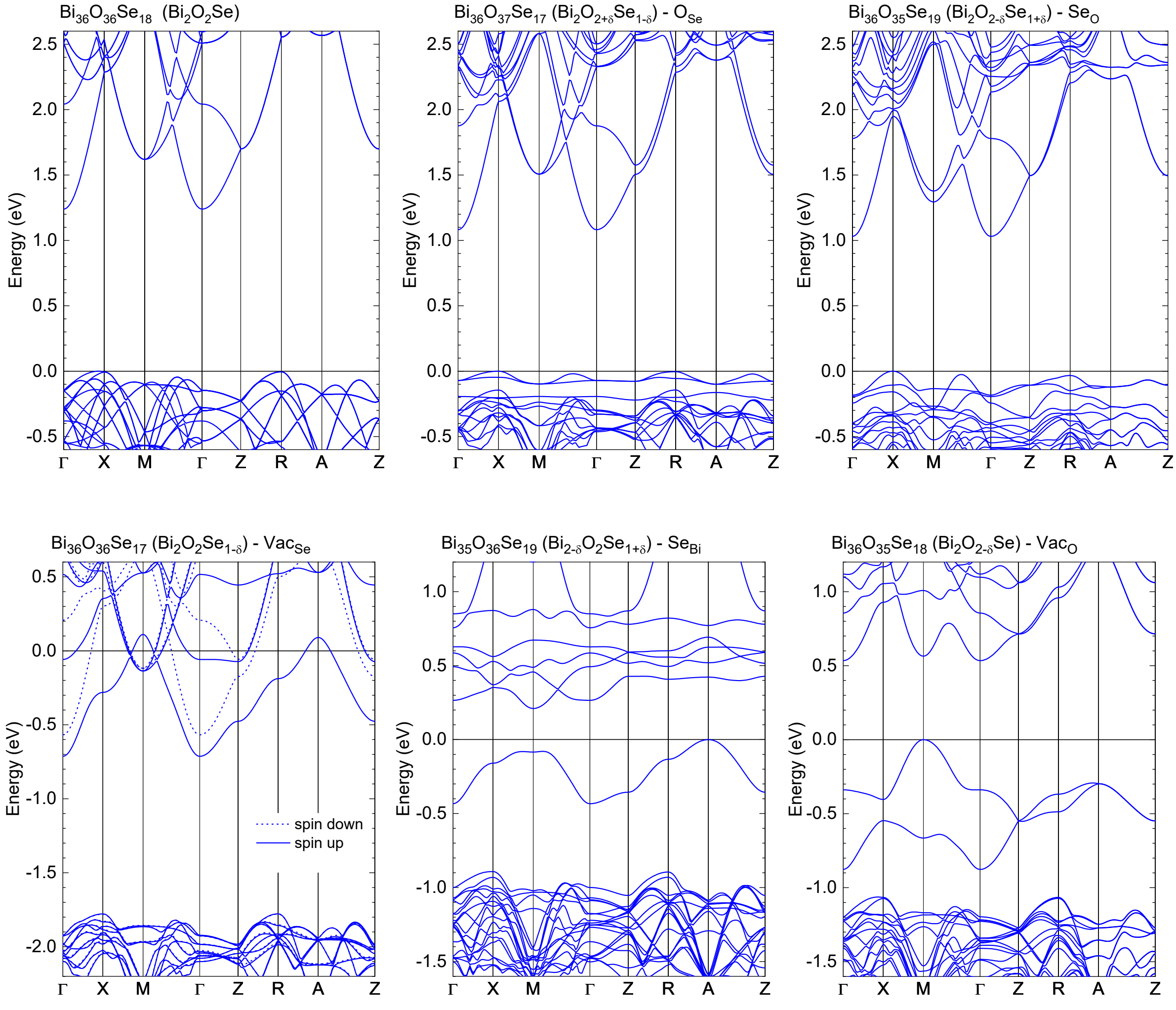

**Figure S8**. Calculated band structure for stoichiometric $Bi_2O_2Se$ and five native defects with the lowest formation energy, using mBJ potential.

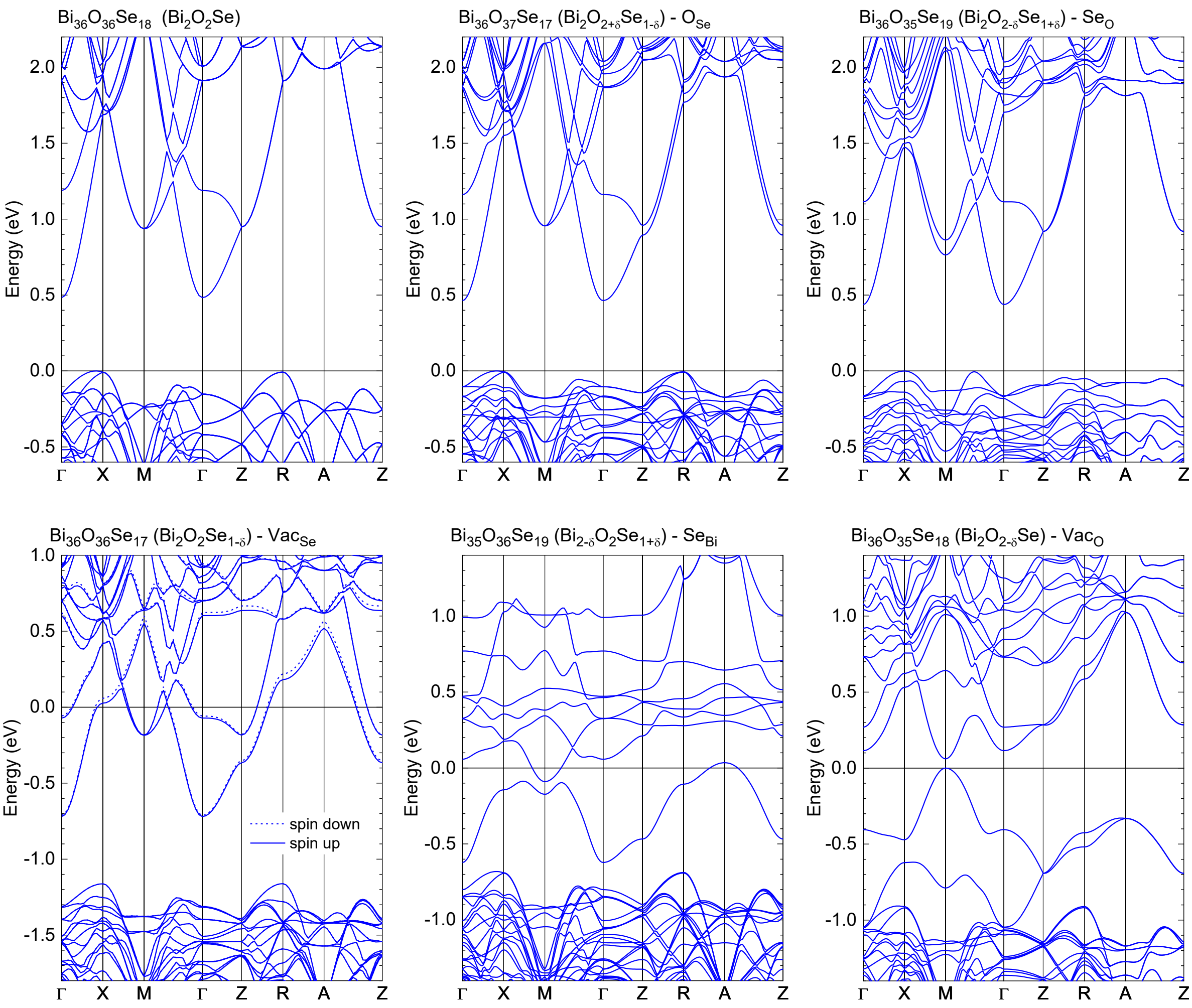

**Figure S9**. Calculated band structure for stoichiometric $Bi_2O_2Se$ and five native defects with the lowest formation energy, using GGA potential.

## 7 Atomic force microscopy

AFM measurements were performed in two modes on freshly cleaved surfaces of single crystals: 1) PeakForce QNM mode, which is used to simultaneously acquire topography and map mechanical properties (evaluated as the logarithm of the DMT modulus) and adhesion (see references 1 and 2) using an AFM Dimension ICON (Bruker); and 2) two-pass Kelvin probe force microscopy (KPFM), which is used to determine topography and surface potential (see references 3 and 4) using an AFM Dimension Nexus (Bruker).

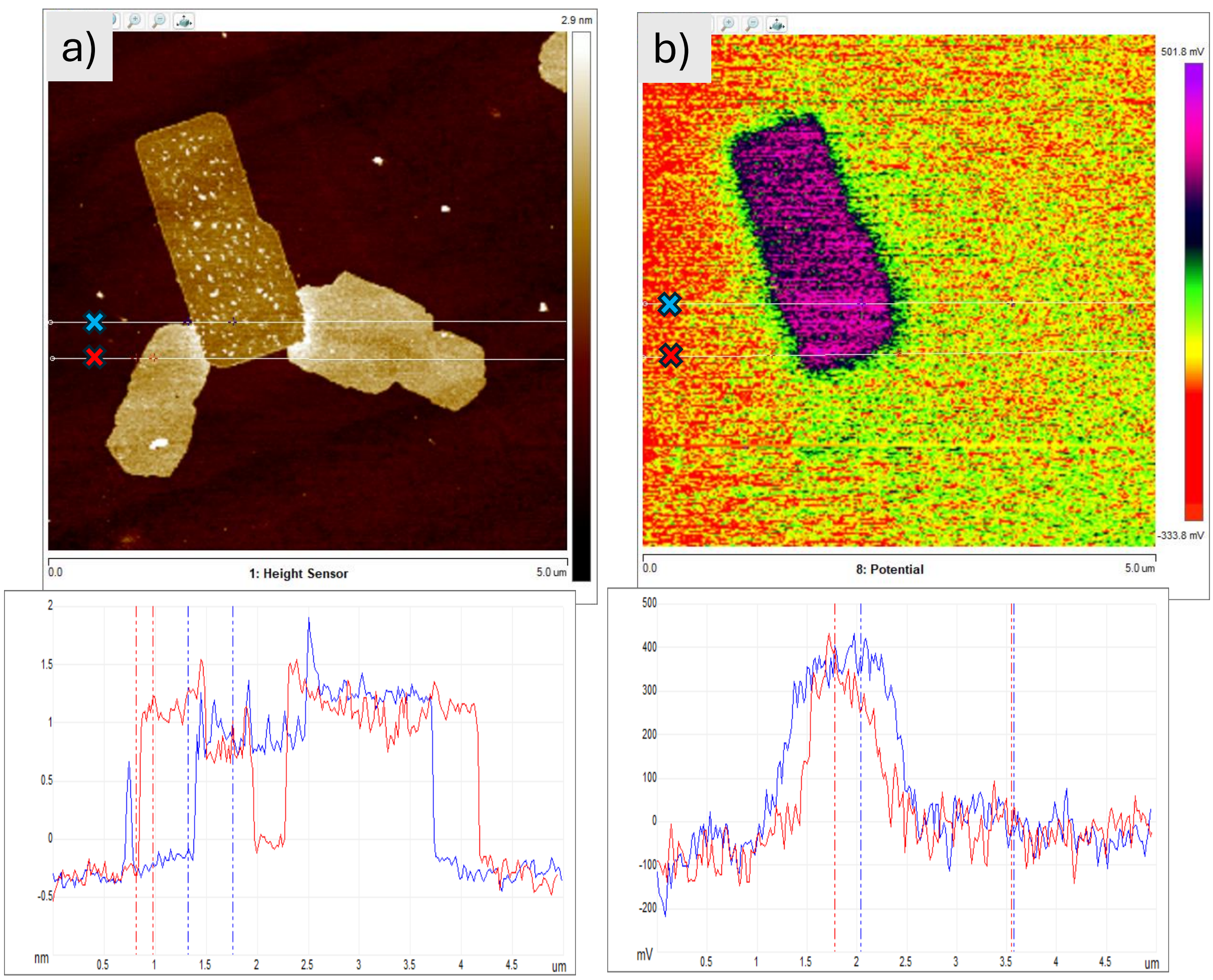

**Figure S10**. AFM derived topology mapping (a) and corresponding surface potential mapping (right) of Sample A. Although both the rectangular and non-rectangular lamellae are topologically visible, only the rectangular lamellae exhibit the surface potential different from the surrounding surface.

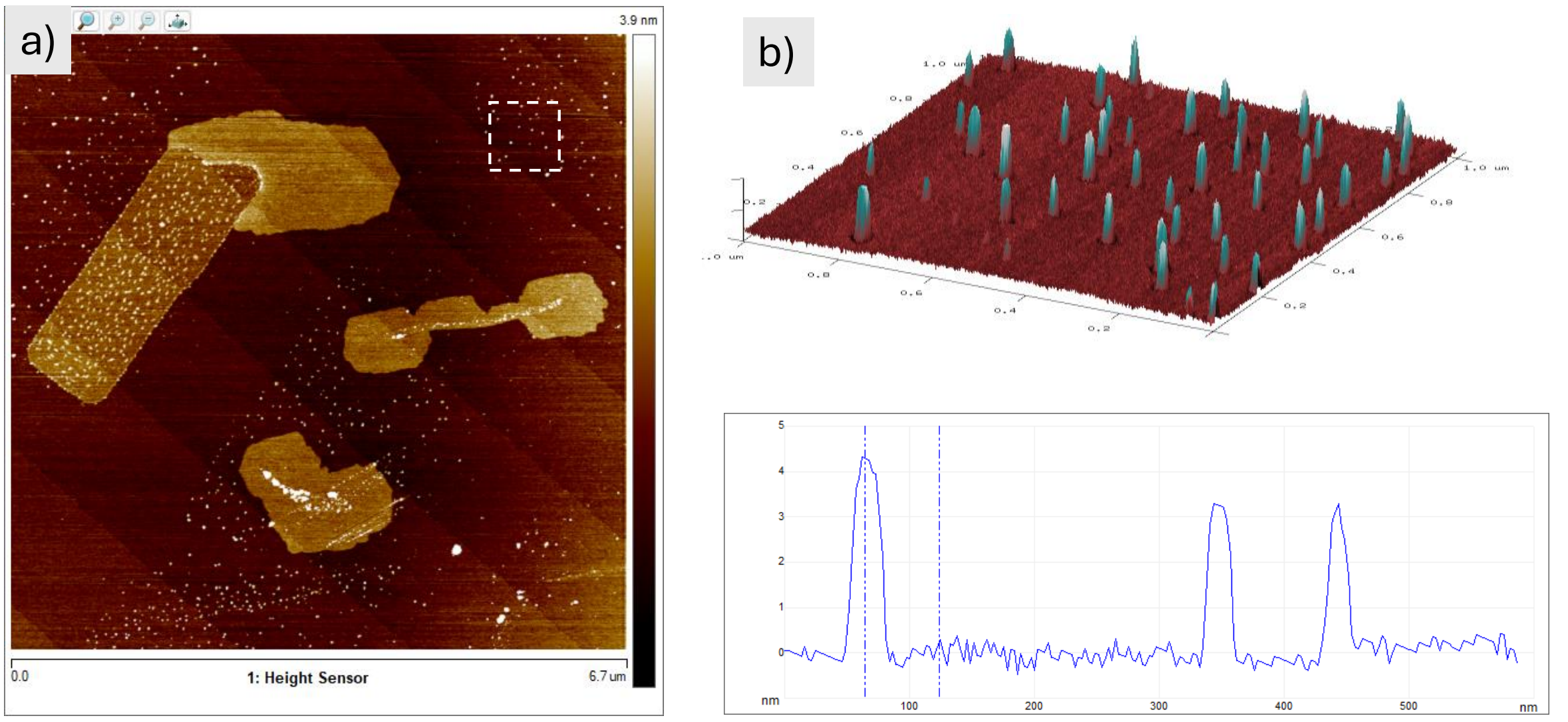

**Figure S11**. AFM derived topology mapping of protrusions in sample A – small white dots (protrusions) in Figure a). The protrusions have heights around 3-4 nm and lateral size 10 -20 nm.

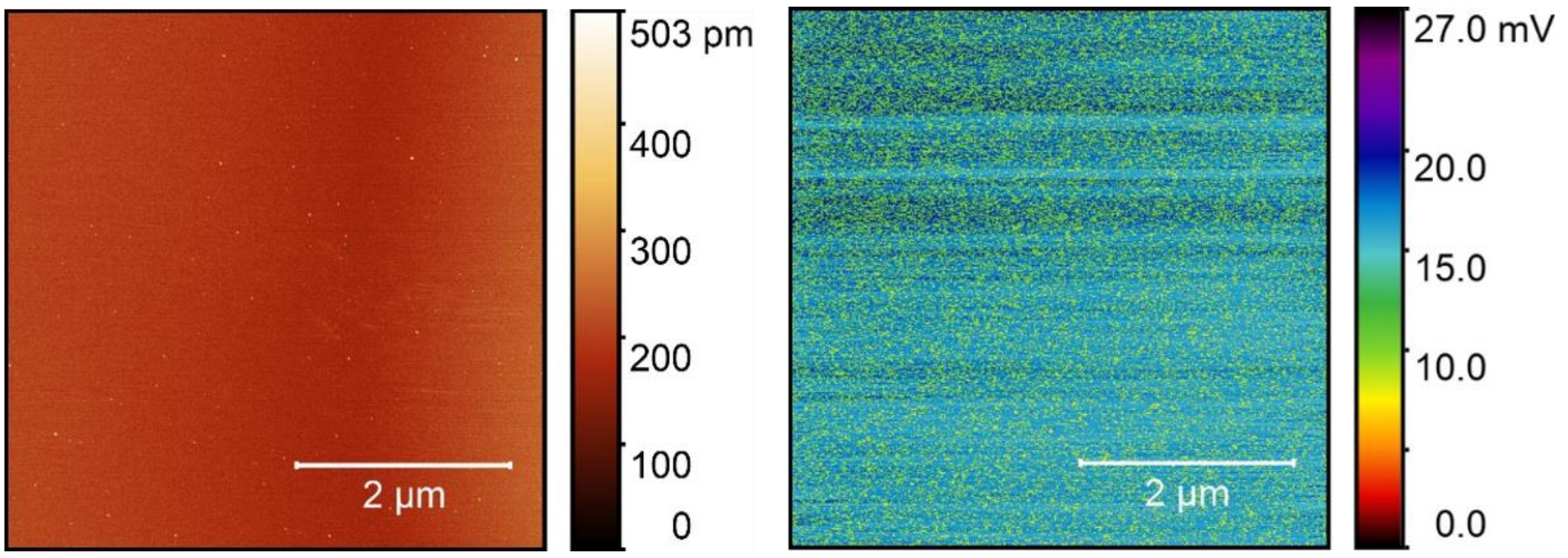

**Figure S12**. Topography (left) and surface potential (right) maps of a completely defect-free region of sample B. These maps supplement Figures 3C and 3F in the main text, showing what was mostly seen.

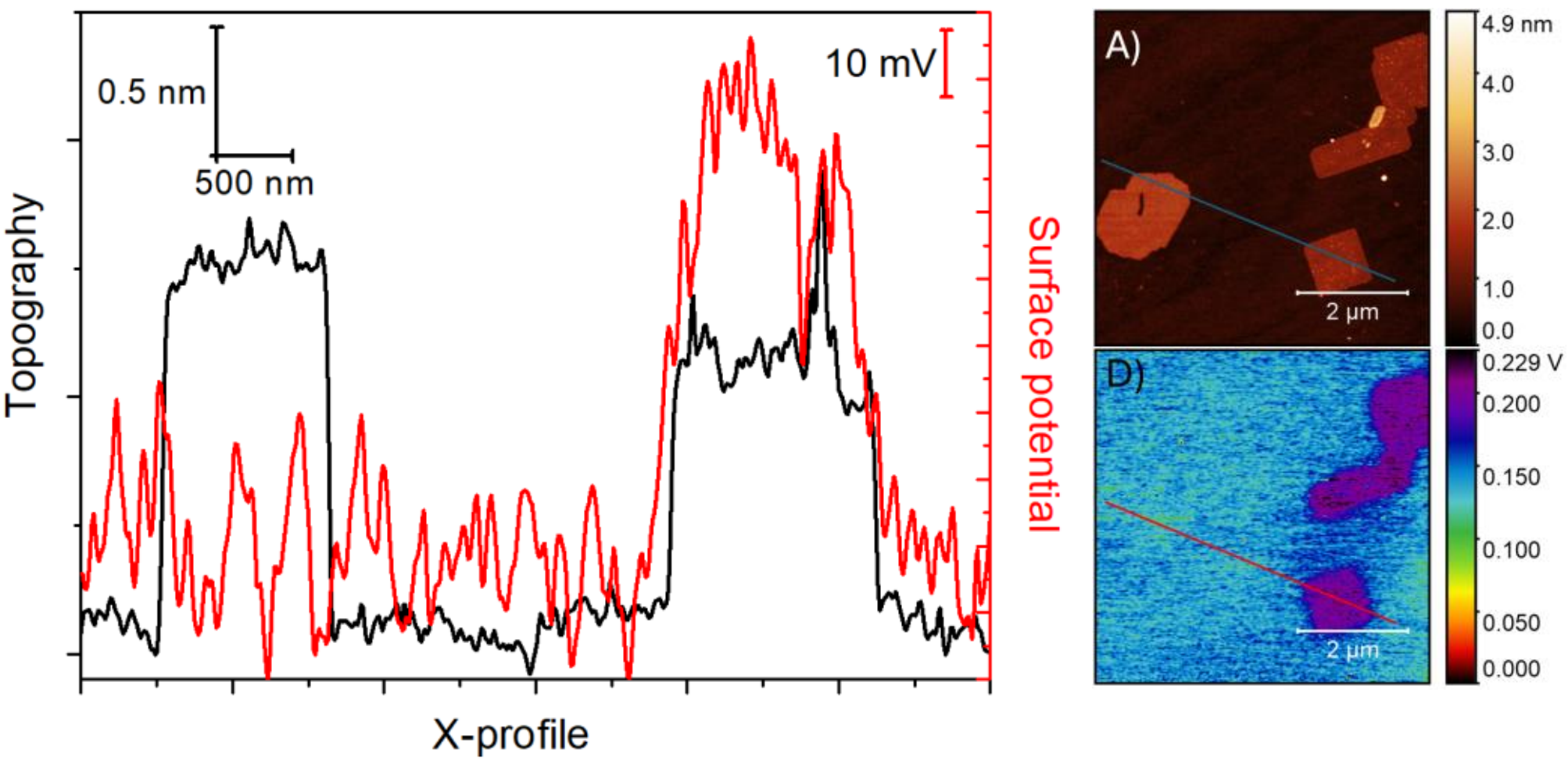

**Figure S13**. (left) Topography and surface potential line profiles corresponding to sample B, taken along the locations indicated by line markers (right). A detailed version of Figure 3 in the main text.

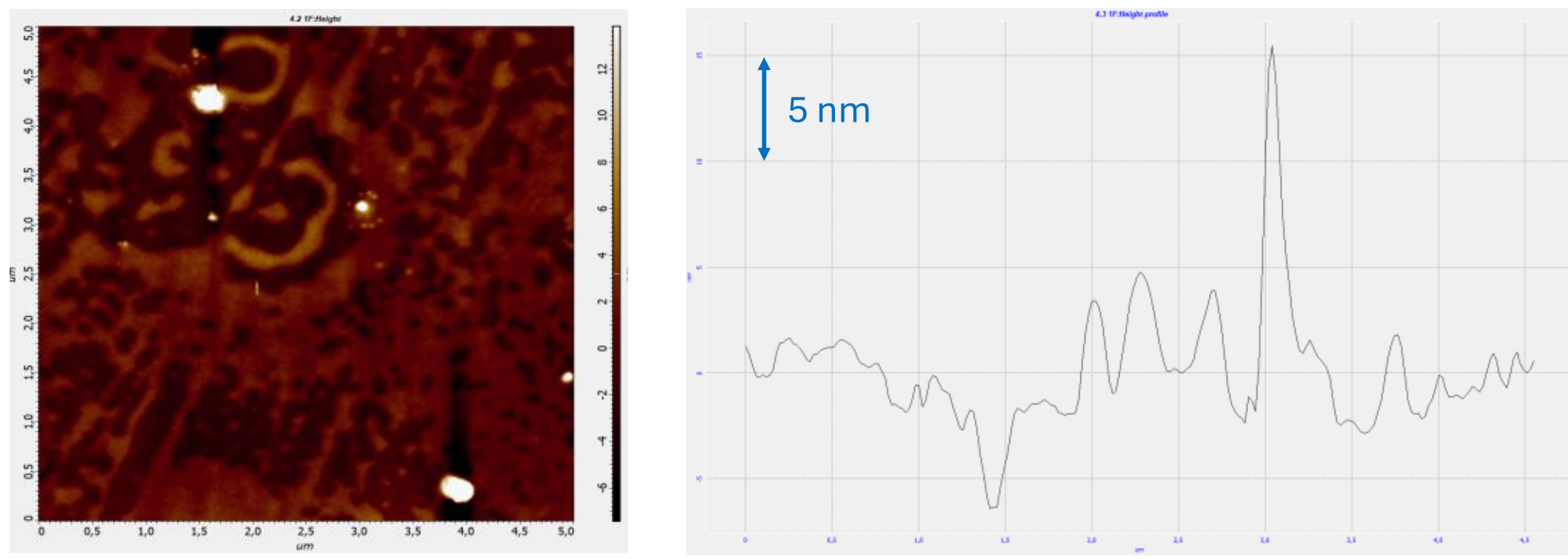

**Figure S14**. (left) Topography (left) and line profile (right) corresponding to sample A stored for 1 year in air. The surface of the sample is subject to destruction due to reactions with oxygen, water, and $CO_2$. The original surface can be restored by removing the surface layer. As can be seen in Fig. S14, the sample loses surface sharpness and becomes covered by degradation products when exposed to air for an extended period. This is most likely due to oxidation, carbonization, and/or hydrolysis. This behavior is consistent with that of other chalcogenide materials, which form a thin oxo-chalcogenide surface layer

over time through the partial substitution of chalcogen atoms by oxygen. This process is accompanied by surface roughening and the formation of protrusions, with oxidation reported to penetrate up to ~14 nm below the surface (see, for example [7].

## 8 Trasport properties of other undoped $Bi_2O_2Se$ samples

This section presents the transport properties of the other undoped samples studied in this research. These results align with the conclusions drawn in the main text. Although such samples are reported in the literature, we were unable to grow samples with large electron concentrations ($10^{19}$ $cm^{-3}$) using the procedures described in the experimental section of the main text. In sample F, the Hall coefficient varies threefold between the low- and high-temperature regions, which indicates significant minority carrier participation. This complicates the analysis of carrier mobility and concentration (see Section 1). This is also evident from the apparent decrease in low-temperature mobility. The low temperature drop in electron concentration in samples E and F can be interpreted as the freezing out of minority holes. The participation of minority carriers in the other samples is less pronounced, to the extent that we consider it negligible. Samples C through F were not comparable to samples A and B in terms of quality and size which prevented a direct comparison. However, they are highly comparable with published data.

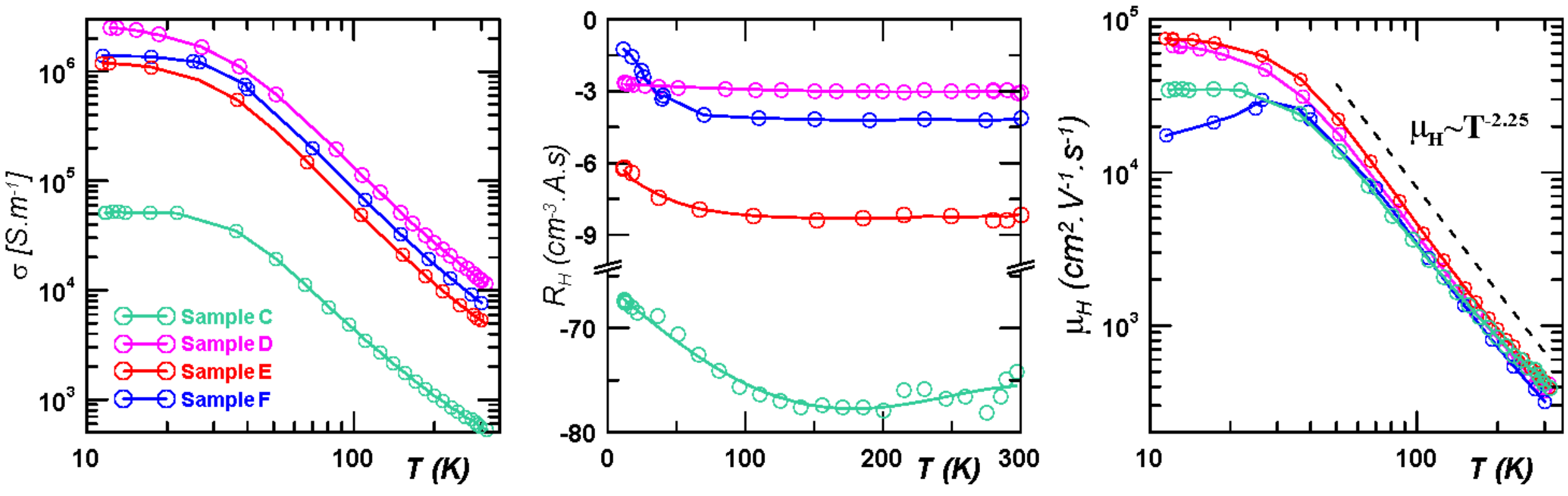

**Figure S15**. The electrical conductivity $\sigma$, the Hall coefficient $R_H$, and the Hall mobility $\mu_H$ for the samples C, D, E and F as a function of temperature.

## 9 Trasport properties of other doped $Bi_2O_2Se$ samples

This section presents the transport properties of other doped samples studied in this research. Regarding the conclusions drawn from Mn doping, the solubility of the doping elements is very small to negligible [8]. However, they largely drive doped materials to very low carrier concentrations, likely due to a shift in stoichiometry. An exception is Erbium doping. We made further attempts to grow crystals with cerium, cobalt, copper, iron, lanthanum, molybdenum, nickel, praseodymium, samarium, tungsten, vanadium, and zinc. However, the crystals were either of low quality or too small, or the resistivity was too high for transport measurements.

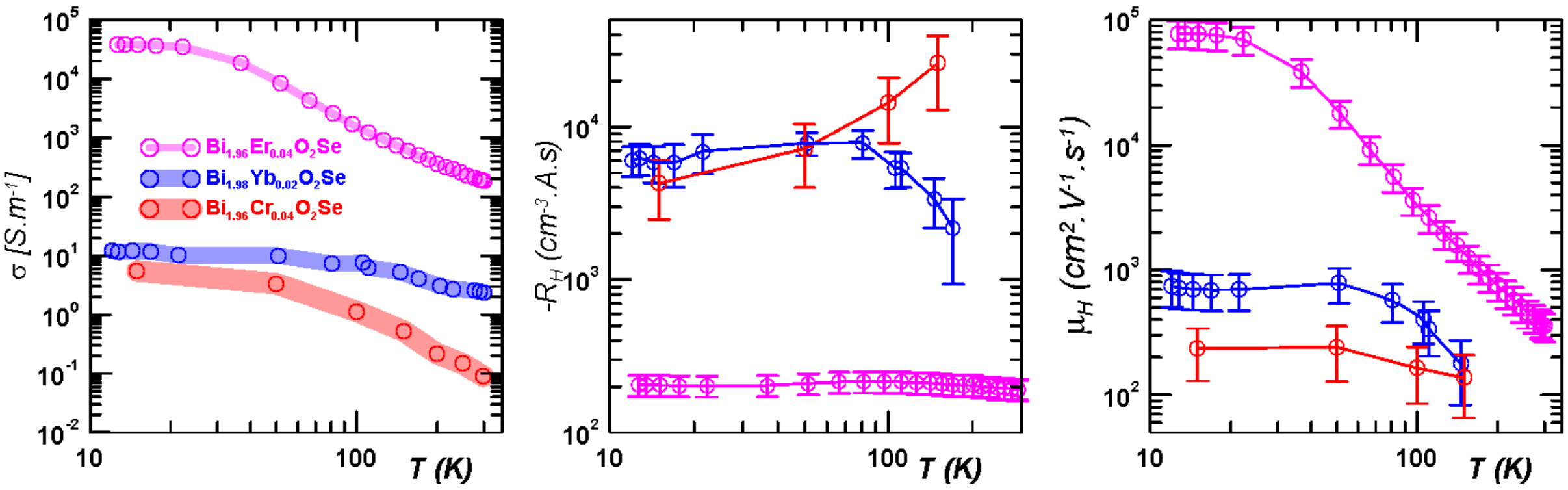

**Figure S16**. The electrical conductivity σ, the Hall coefficient $R_H$, and the Hall mobility $\mu_H$ for the doped samples $Bi_{2-x}M_xO_2Se$ (M=Cr, Yb, Er).

## Acknowledgements

This work was supported by the Czech Science Foundation (Project No. 22–05919S). This work was also supported by the Project TERAFIT - CZ.02.01.01/00/22_008/0004594 cofinanced by the European Union and the Ministry of Education, Youth, and Sports of the Czech Republic, and by the grant of the Ministry of Education, Youth, and Sports of the Czech Republic (Grant No. LM2023037). We are grateful to Vít Novák for supplying the GaAs sample. We gratefully acknowledge Dr. Hartmut Stadler (Bruker Nano Surfaces & Metrology, Karlsruhe, Germany) for performing the KPFM measurements.